\documentclass[12pt,preprint]{aastex}

\begin{document}

\title{Mini High Velocity Clouds}

\author{G. Lyle Hoffman}
\affil{Dept. of Physics, Lafayette College, Easton, PA  18042; 
hoffmang@lafayette.edu}

\author{E.E. Salpeter}
\affil{Center for Radiophysics and Space Research, Cornell University, Ithaca, 
NY  14853; ees12@cornell.edu}

\and

\author{Ajay Hirani}
\affil{Dept. of Physics, Lafayette College, Easton, PA  18042}

\received{}
\accepted{}
\slugcomment{To appear in Astronomical Journal}

\begin{abstract}

Neutral hydrogen mapping of several small, very low column density 
High Velocity Clouds (HVC), using the Arecibo
telescope, is reported.
Some were found serendipitously superimposed at distinct velocities on larger 
HVCs; others were found to comprise the four most isolated low column density
sources in the observations of Lockman et al. but proved to be individually
much smaller than the Green Bank 140 Foot telescope beam.
We call these clouds ``mini-HVC'' to distinguish them from the larger and denser 
Compact High Velocity Clouds (CHVC).
Peak column densities are typically a few $\times 10^{18}~{\rm cm}^{-2}$ averaged
over the Arecibo $3\farcm2$ beam, and diameters to our detection limit $\sim 10^{18}~{\rm cm}^{-2}$
range from $9\arcmin$ to $35\arcmin$.
These column densities and angular diameters overlap with those for CHVC, but are typically smaller.
We consider three possiblities:  (1)  that most mini-HVC are related to the Magellanic Stream,
(2) that most mini-HVC are part of M31's retinue of CHVC, or (3) that the mini-HVC are simply the
low column density tail of the distribution of CHVC.
None of these possibilities can be rejected as yet, given the selection biases in our sample.
We also discuss controversies about the amount of ionized hydrogen in CHVC and mini-HVC, 
which may be mainly ionized, 
and the implications of these small clouds for Lyman Limit Systems.

\end{abstract}


\section{Introduction}

The major High Velocity Cloud (HVC) complexes have appreciable angular sizes 
and are thought to lie mainly closer than the Large Magellanic Cloud
(or, in the case of the Magellanic Stream, up to that distance),
closer than most dwarf irregular galaxies (dIrr) in the Local Group.
Compact High Velocity Clouds (CHVC) were first identified by \citet{BB99}
as HVC that are small ($\lesssim 1\arcdeg$) in angular extent and
significantly removed from any of the major HVC complexes.
Subsequent work by many researchers has shown that CHVC have lower 
metallicity than the major complexes \citep{SGFP02}, kinematics that appear to center on 
the barycenter of the Local Group rather than the Milky Way \citep{dHBB02b}, and internal 
structure consistent with their being supported by a combination of dark 
matter gravity and external pressure \citep{BBC01, dHBB02a, SMW02}.
Several lines of evidence point to distances of order 50-200 kpc for CHVC
\citep{BBdH02, MP03, PBV+03}, typical of many of the Local Group dIrr.
Individual CHVC have central gas density apparently below the threshold 
for star formation, and careful searches for stars in several CHVC have 
found none \citep{SB02, DSD+02, HSK03}.
CHVC may be good candidates for infall of (almost) primordial material into disk galaxies
\citep{BSTHB99, BB99}, which would replenish hydrogen and provide some 
dynamical stirring to encourage star formation.
This could be of interest for large and small disk galaxies, but more so for small dwarf irregular 
galaxies (dIrr) because of their larger ratio of surface area to volume.
Two possible examples of such CHVC/dIrr mergers have been reported \citep{RSB02, HBSC03}.

We have previously reported Arecibo\footnote{The 
Arecibo Observatory is part of the 
National Astronomy and Ionosphere Center, 
which is operated by Cornell University under a management agreement 
with the National Science Foundation.} 
mapping of one CHVC and one small HVC that did not quite meet the
size and isolation requirements to be labelled a CHVC \citep{HSP02}.
\citet{BBC01} show Arecibo mapping of several more CHVC to similar 
sensitivity.
While the main purpose of those maps was to study the \ion{H}{1}/\ion{H}{2}
ionization balance, a major surprise was our discovery of a small cloud
of \ion{H}{1} superimposed on WvW~491, differing in velocity by 
66~km~${\rm s}^{-1}$ which is more than twice the FWHM of the \ion{H}{1}
profile of either WvW~491 or the cloudlet.
We have dubbed such cloudlets ``mini-HVCs.''
Preliminary mapping of WvW~479 revealed two more such mini-HVC and
led us to speculate that many of the low column density \ion{H}{1}
sources --- at least those well removed from major HVC complexes --- 
found by \citet{LMPU02} (hereafter refered to as LMPU) using the 
140 Foot telescope at Green Bank
might be similar cloudlets, considerably smaller than the 21\arcmin~ beam
of that instrument.
We undertook to map the four most isolated LMPU sources, those that
appeared in their central beam but not in the preceding nor trailing
reference beams.
In this paper we report our mapping of these four LMPU fields along
with the mapping of WvW~479 and WvW~413 (the latter apparently devoid
of mini-HVCs).

Our Arecibo observing procedures are detailed in Sect. 2, with 
results presented in Sect. 3. 
In Sect. 4, we discuss the relationship among mini-HVC, CHVC and 
the major complexes (the Magellanic Stream in particular),
consider photoionization mechanisms, explore the
column density-diameter relationship for mini-HVC and CHVC, and
ponder the possible connection between these systems and Lyman Limit
Systems.
Conclusions and a summary follow in Sect. 5.

\section{Arecibo Observations}

Preliminary mapping of WvW 413 and WvW 479 in driftscan mode was conducted
in 2001 June-July.
Using 16 driftscans separated in declination by $3\farcm2$ and each 480 s
long, we mapped an area about $109\arcmin$ by $51\arcmin$.
In the allocated time we were able to complete 6-10 repetitions
of each declination strip which we averaged to achieve rms noise about 
5 mJy in velocity channels 0.65~km~${\rm s}^{-1}$ wide.
For features having FWHM about 20~km~${\rm s}^{-1}$, this gave a minimum
detectable flux integral of order 0.2~Jy~km~${\rm s}^{-1}$, sufficient
to detect the core of each CHVC and any mini-HVC with central column density
above a few~$\times 10^{18}$~${\rm cm}^{-2}$.

Follow-up mapping of WvW 413 and WvW 479 was conducted in 2002 June-August,
along with mapping of the four LMPU sources.
To achieve greater sensitivity, we used position-switched mapping with
longer integration times -- 5-40 minutes on-source (with equal time on
each reference beam) per point on sparsely sampled maps.
We used 6.25 MHz bandwidth (0.65~km~${\rm s}^{-1}$ channels) and 9-level 
sampling.
This produced rms noise 0.8-2 mJy, depending on integration time and
zenith angle, and secure detections for narrow line features as low as
$8 \times 10^{17}$~${\rm cm}^{-2}$.
A noise diode was fired after each on-off pair.
Calibration was provided by the NAIC staff.

WvW~413, for which preliminary mapping did not reveal any superimposed 
mini-HVCs, was mapped along two orthogonal diameters aligned with
the cardinal Galactic directions.
Points were spaced by one beamwidth, as shown in Fig. 1.
Integration times varied from 5 minutes on-source at the core of the 
HVC to 20 minutes on-source at the outermost points.

Preliminary mapping for WvW~479 exposed two superimposed mini-HVCs.
In follow-up work, we obtained longer integrations on the positions
found in the preliminary map to have the highest column density for each,
along the line joining those two centers both toward and away from
the other mini-HVC center, with beams spaced by a beamwidth, and 
at four additional points around each center to complete a hexagon
at beamwidth spacing from the center point, as shown in Fig. 5.

Each LMPU source was first mapped in an hexagonal pattern covering the
140 Foot beam with points spaced two Arecibo beamwidths apart.
Integration times were limited to 5 minutes on-source for the first
pass.
Marginal detections were followed-up with longer integration times,
up to 20 minutes on-source, giving a typical 3-sigma detection limit
of about $4 \times 10^{17}$~${\rm cm}^{-2}$ assuming a profile
width of $\sim 25$~km~${\rm s}^{-1}$.
Once a clear detection was achieved, we mapped interstitial points to
determine the highest column density reached in the core of the mini-HVC
and continued to surrounding points to assess the full extent of the
cloud.
In the event that emission was detected in a reference beam, we
confirmed that detection with a fresh on-off pair (the on beam
in this case being centered on the putative detection) and then
mapped around that position in a hexagonal pattern with beamwidth 
spacing just as if it had been detected within the LMPU field.
The full complement of observed points is recorded in Fig. 7.

\section{Results}

Results from the Arecibo mapping are summarized in Table 1 and Figs. 1-8.
The columns of Table 1 give, respectively, each object's name, R.A.
and Dec. (J2000), the peak observed flux, approximate major and minor
diameters to the outermost detected point with an uncertainty of order
half the beamwidth (i.e., $2\arcmin$; $>$ indicates that our observations
do not reach the detection limit on at least one side), 
an approximate position angle for the major axis, the peak column density,
the systemic velocity (midpoint between velocities at which the spectrum
from that central position falls to half its peak value), and the
velocity width between those 50\% points (i.e., the FWHM).
In the subsections that follow, we elaborate on individual sources.

\subsection{WvW 413}

The drift scan mapping of the HVC WvW 413 disclosed a well-defined, 
approximately elliptical core centered near 
$22^{h}19^{m}15.1^{s} + 25\arcdeg32\arcmin00\arcsec$ (B1950) which translate
to the J2000 coordinates listed in Table 1.
The outskirts appeared irregular, but the sensitivity was not high enough
to determine whether the irregularity was intrinsic or due to noise.
For follow-up studies, we chose to map along axes aligned with Galactic
N-S and Galactic E-W to investigate whether the scalelengths of the
\ion{H}{1} emission were approximately equal in all directions
as expected if the truncation is mainly due to photoionization by
extragalactic UV sources, or more abrupt toward Galactic North as
would be the case if the HVC (which resides in the southern Galactic
hemisphere) is within a few kpc of the Milky Way's
disk and truncation is due to photoionization by Galactic sources or
to collisions with hot halo gas.

The results of the follow-up mapping are shown in Figs. 1-4.
Fig. 1 presents a map of the observed points, with numbers
indicating observed \ion{H}{1} column density in units of 
$10^{18}$~${\rm cm}^{-2}$ and exes indicating observed positions with no
detected emission.
Spectra obtained from the center beam and from beams near the outer 
ends of each diameter are shown in Fig. 2.
Column density as a function of radius for each of the four cardinal 
Galactic directions is shown in Fig. 3, and the ``rotation curve''
of the HVC, a plot of systemic velocity vs. radius along both diameters, 
is shown in Fig. 4.
The variation in velocity along the E-W axis is probably due to noise,
but there does appear to be a systematic trend to the systemic velocity
along the N-S axis, perhaps indicating rotation.
The amplitude is less than the width of a typical profile, however.

The column density runs for Galactic north and west in Fig. 3 have 
distinctly steeper slopes than those for South and East.
Since the HVC has Galactic coordinates $l = 85\arcdeg$, $b = -26\arcdeg$,
Galactic North is toward the plane of the Milky Way and Galactic West
is toward its center.
This could be an indication that Galactic sources contribute a significant
fraction of the ionizing UV, or that the HVC is moving in that direction
relative to a hot external medium.
The former would require that the HVC lie within a few kpc of the disk 
so that the UV flux from Galactic sources is not too weak.
The latter requires either that the HVC lie within the Galaxy's corona
(and have space velocity toward Galactic NW) or that the hot gas extends
far outside the Milky Way.

\subsection{WvW 479 and Superimposed Mini-HVCs}

The drift scan mapping of the HVC WvW~479 disclosed a well-defined, 
approximately elliptical core centered near 
$01^{h}21^{m}44.8^{s} + 25\arcdeg24\arcmin24\arcsec$ (B1950) which translate
to the J2000 coordinates listed in Table 1.
Two mini-HVCs, one south and the other SE of the HVC's core, were also 
apparent in the driftscan map.
We have labeled these WvW~479 min1 and min2, respectively, even though
they may be chance superpositions.
The two mini-HVCs had nearly identical velocities, $-248$ and $-245$ 
respectively, so for follow-up studies we chose to map a one-beamwidth
spaced hexagon around each core to verify that we had identified the
maximum column density position in each, and along the line joining 
the two to determine whether or not they were connected at column
densities below the sensitivity of the driftscan map.

The results of the follow-up mapping are shown in Figs. 5-6.
Fig. 5 presents a map of the observed points with numbers
indicating observed \ion{H}{1} column density in units of 
$10^{18}$~${\rm cm}^{-2}$ and exes indicating non-detections.
Spectra obtained from the centers of the two mini-HVCs are shown in Fig. 6.
While the allocated time was not sufficient for us to map the entire
distance between the two mini-HVCs, min1 was no longer detected
in a 30 minute (on-source) scan at a point $6\farcm4$ east of center.
Along the westward track from min2, the column density fell smoothly
to $8.4 \times 10^{17}$~${\rm cm}^{-2}$ at the point $9\farcm6$ from
the core, but the next point outward unexpectedly rebounded to
$2.5 \times 10^{18}$~${\rm cm}^{-2}$.
We ran out of telescope time at that point and so do not know
how much further the mini-HVC extends.
It was not detectable in the draftscan mapping beyond that point,
so we know the column density does not rise back above 
$\sim 3 \times 10^{18}$~${\rm cm}^{-2}$ before min1 is reached.

\subsection{LMPU Sources}

\subsubsection{LMPU025}

Four mini-HVCs were found in and around the 140 Foot beam area for
LMPU025, and two more were found in the reference beams $6^{m}$ to
the east.
The Arecibo beam positions for those in the 140 Foot beam area
are shown in Figs. 7a-c, and those in the area of the
reference beams are shown in Fig. 7d.
Mini-HVC LMPU025-163, at velocity $\sim -163$~km~${\rm s}^{-1}$,
was found to span the southern third of the 140 Foot beam area and
to extend a few arcmin beyond to the NE and SW as shown in Fig. 7a.
The highest column density point (Fig. 8a) falls in the 
SE portion of the 140 Foot beam area, but the core of the mini-HVC 
is not well-defined.
The signal-to-noise ratio in individual spectra is too low for us to
determine if there are any systematic trends in systemic velocity
from one end of the cloud to the other.

It may be possible that LMPU025-291 and LMPU025-335 are part of a 
single system, but the 44~km~${\rm s}^{-1}$ difference in systemic
velocity and abrupt drop in systemic velocity from 
$\sim -330$~km~${\rm s}^{-1}$ for the points near ($-0\fdg10$,$29\fdg3$)
in the map to $\sim -290$~km~${\rm s}^{-1}$ for the points
at R.A. offset $-0\fdg16$ makes that unlikely.
Within each of the two there is no systematic trend in velocities.
Spectra of the highest column density points of both mini-HVC are shown in 
the Fig. 8a.

Traces of emission at $\sim -396$~km~${\rm s}^{-1}$ were evident in a
couple of the westernmost points in the 140 Foot beam area of LMPU025,
so we traced the emission further west to disclose an unusually
high column density mini-HVC lying just outside the 140 Foot area
as shown in Fig. 7c.
The spectrum obtained from the highest column density point is shown 
in Fig. 8a.
There is no systematic trend to the velocities for most of this
mini-HVC, but the two easternmost points have velocities 
$\sim 5$~km~${\rm s}^{-1}$ higher.

The allocated time did not permit a complete mapping of the two
mini-HVCs found in the reference beam area, but we were able to locate
the position of highest column density for each.
The diameters listed in Table 1 are therefore more approximate ---
lower bounds, perhaps --- than those for the mini-HVCs we mapped
more completely.
The spectra obtained from the highest column density points of each
are shown in Fig. 8b.
The close proximity in velocity of these two mini-HVC to two of those
found in the main 140 Foot beam area suggests that perhaps they are
connected by bridges of more completely ionized gas.

\subsubsection{LMPU236}

The single mini-HVC found in the 140 Foot beam area LMPU236 is the only
mini-HVC detected in the northern Galactic hemisphere to date.
It also is the only known positive velocity mini-HVC and has the 
lowest peak column density of all the mini-HVCs found so far.
The map of observed points is shown in Fig. 7e; detectable emission
spans only one-third of the 140 Foot beam area.
The spectrum from the highest column density point is shown in Fig. 8c.

\subsubsection{LMPU369}

A single mini-HVC was found in the 140 Foot beam area of LMPU369, 
with its peak column density just outside the area as shown in
Fig. 7f.
The spectrum from the peak column density point is shown in Fig. 8d.
However, another mini-HVC was found in the reference beam, $6^{m}$ in R.A.
($1\fdg5$ of arc) east of the northermost point in the LMPU369 area.
The peak column density point was found to be one Arecibo beamwidth 
further north.
Points with detectable emission are shown in Fig. 7g,
and the spectrum from the highest column density point is shown
in Fig. 8d.

\subsubsection{LMPU387}

Three mini-HVCs in addition to the northern end of the Magellanic Stream
were found in the 140 Foot beam area of LMPU 387.
LMPU387-230 and LMPU-297 are almost superimposed on one another in the
SE edge of the area, as shown in Fig. 7h and 7i, but spectra (Fig. 8e) from those 
points show the two at clearly distinct velocities.
LMPU387-275 lies to the NE of the 140 Foot beam area.
All three have fairly well-defined edges, fill only one-third to 
one-half the 140 Foot beam area, and overlap its edge.

\subsection{Statistics of Individual and Peak Column Densities}

Not counting the CHVCs WvW 413 and WvW 479 themselves, we have detected
flux in 229 individual locations spanning 14 distinct mini-HVCs.
The 14 peak column densities are given in Table 1 and Fig. 9, and
the distribution function for the 229 individual column densities 
is shown in Fig. 10.
For $N_{HI} \gtrsim 2 \times 10^{18}$~${\rm cm}^{-2}$ the distribution
function is roughly proportional to ${N_{HI}}^{-1}$, somewhat similar 
to the distribution function of column densities for Lyman Limit Systems
(see also Sect. 4.4).
Detections in the lowest bin are highly incomplete, but few detections 
should be missing in the third bin ($1$ to $2 \times 10^{18}$~${\rm cm}^{-2}$).
Thus the distribution function changes slope below 
$2 \times 10^{18}$~${\rm cm}^{-2}$, partly because the ratio of ionized to neutral hydrogen increases there (see also Sect. 4.3).
For peak column densities of the mini-HVCs, there are only 3 with
$N_{HI}$ in $(2$ to $4) \times 10^{18}$~${\rm cm}^{-2}$ compared with 7 in
$(4$ to $8) \times 10^{18}$~${\rm cm}^{-2}$ (see also Fig. 11).

Our observations of mini-HVCs were appreciably undersampled by the Arecibo beam,
so the true peak $N_{HI}$ could be larger than the observed peaks in Table 1 and
Fig. 9.
However, the typical true/observed ratios depend on how large flux 
fluctuations are in the vicinities of the peak locations.
As can be seen in Figs. 5 and 7, these fluctuations are surprisingly small.
For instance, there are practically no non-detections near the observed peaks.
We have also looked at how many individual fluxes in each mini-HVC are above 
$3/4$ of the observed peak flux.
The average number is 4.2.
It is therefore unlikely that the true/observed flux ratio is above $4/3$
in very many cases.
The near-constancy of the flux in the central regions is somewhat surprising,
given the irregular shapes of the outermost regions and the rapid fall-off 
there.

\section{Discussion}

\subsection{Distribution of Mini-HVCs on the Sky}

Of the 15 mini-HVCs we have found to date, only one (LMPU236+242)
is in the northern Galactic hemisphere.
The three LMPU387 mini-HVCs are all superimposed on the northern
extension of the Magellanic Stream (MS), but at velocities distinct from
the MS at that point on the sky.
The LMPU369 and LMPU369 OFF mini-HVCs are several degrees west of
the MS, and the LMPU025 mini-HVCs along with those superimposed on
WvW~491 and WvW~479 are spatially in the vicinity of the recently reported
northern extension of the MS \citep{BT04,SDKB02,LMPU02}, but at distinct velocities.

The moderate proximity of all but one of the as-yet-discovered
mini-HVCs to the MS raises the question:  Are most of the mini-HVC
somehow associated with the MS in spite of the velocity difference?
A bifurcation of the MS has already been noted by \citet{SDKB02},
although with a smaller difference in velocity.

A second possibility is that some of the mini-HVCs are the low column density 
tail of the retinue of CHVCs which surround M31 \citep{TBW+04} and M33
\citep{BT04}, if that
retinue extends about 200 kpc from M31 or M33.
The physical \ion{H}{1} diameters of mini-HVC would in that case 
be comparable to those of typical CHVCs since the mini-HVCs would
be 3-4 times more distant.
It would be necessary that there be some beam dilution of the
cores of the mini-HVCs unless the M31 population of CHVCs has
inherently lower column density cores than the Milky Way
population.
However, the cores of typical CHVCs span several Arecibo beams,
which implies that beam dilution would be minimal if they were
moved 3-4 times further away, and for our mini-HVC we observe similar
column densities spanning 2-3 Arecibo beams around the maximum observed
column density, suggesting that they have not suffered much beam
dilution.
Mapping, with adequate sensitivity, of the sky in the vicinity
of M31 (as close as the northern limit of the Arecibo declination
range permits) using the Arecibo L-band Feed Array should help
distinguish between MS and M31/M33 affiliation.
Telescopes with larger beams would suffer beam dilution of the
mini-HVC cores, mandating longer integration times.

A third possibility is that the mini-HVCs have nothing to do 
with the MS on the one hand, nor with M31 or M33 on the other, 
and are just the low column density tail
of the Milky Way's CHVC population.
This possibility becomes more attractive with the additional conjecture 
that distance from the Milky Way tends to increase with decreasing
central column density $N_{HI}$.
Typical CHVC distances might then be only 50 to 100 kpc and mini-HVC 
distances typically 200 kpc with little difference in absolute linear sizes.
The concentration of the mini-HVCs discovered to date toward
the MS and M31 would then just be a selection effect, a 
consequence of our having chosen HVCs mainly in that direction
for edge mapping and of small number statistics in the LMPU
isolated central beam sources.
A prediction for very sensitive \ion{H}{1} mapping from 50 to 200 kpc
away from M31 and M33 is decreasing $N_{HI}$ but little change in 
angular size.

\subsection{Column Density-Diameter Relationship}

The relation of peak column density $N_{HI}$ to diameter $D_{HI}$ 
for our 15 mini-HVCs is shown in the left-hand panel of Fig. 9
and compared to that for CHVC in the right-hand panel.
Here we have simply taken the peak column density as recorded by
whatever telescope was used.
CHVC mapped only with Parkes or Dwingeloo may have central column densities
reduced by beam dilution by comparison to those mapped at Arecibo.
The diameters are also heterogeneous; most of the CHVC were not mapped
to as low a column density as the few mapped at Arecibo.
We have simply taken the diameter to the outermost point at which flux was
detected.
Therefore many of the HIPASS and LDS data points would move to the right
and upward in the figure if those CHVC were remapped at Arecibo.
For the mini-HVCs and CHVCs separately there is some trend toward 
lower column density for smaller angular diameter, but with a large scatter.
According to the conjecture mentioned above (third possibility, Sect. 4.1),
the small angular size may be due to larger distance rather than 
smaller linear size.

\subsection{Photoionization Controversies}

For hydrogen gas with volume density $n_{H}$, ionized at rate $I$ by
extragalactic UV plus collisions, there is a critical value $N_{crit}$
of \ion{H}{1} column density above which there is little \ion{H}{2}
\citep{CB02, CSB01, SMW02}.
Below $N_{crit}$ the neutral component decreases very rapidly with
decreasing total hydrogen column density $N_{H, tot}$ so the \ion{H}{1}
scalelength is a few times smaller than the scalelength for total hydrogen.
The value of $N_{crit}$ is proportional to the ratio $I / n_{H}$.
If $N_{crit}$ were $<< 10^{19}$~${\rm cm}^{-2}$ for a CHVC, 
the ionized hydrogen would contribute little and the total gas mass
(assuming 200 kpc distance) would be about $10^{6}$~${\rm M}_{\sun}$.
If $N_{crit} \sim (1-2) \times 10^{19}$~${\rm cm}^{-2}$ as for
spiral galaxies, the true total hydrogen scalelength is much larger,
and $M_{H, tot}$ is $10^{7}$~${\rm M}_{\sun}$ or more.
Similar reasoning must apply to the mini-HVC, although (if they lie
in a similar ionization field) their smaller central neutral column
density would imply an even larger fraction of ionized gas.

Observations of the outer disks of spiral galaxies suggest
$N_{crit} \sim 2 \times 10^{19}$~${\rm cm}^{-2}$, as does our study
of the dwarf irregular DDO154 \citep{HSC01}.
This agrees with the theoretical expectations if ionization is due solely
to extragalactic UV.
Similar values of $N_{crit}$ are suggested for CHVC at 200 kpc distance
\citep{SMW02}.
However, the recent detection of H$\alpha$ in several CHVC by 
\citet{TWMHR02} and \citet{PBV+03} exceeds that which could be produced solely by
photoionization by extragalactic UV;
if CHVC are outside the Milky Way halo they must be ionized collisionally
by a hot intragroup medium \citep{BB02}, which increases $I$ 
appreciably.
Highly ionized gas has been found around the major HVC Complex C \citep{FSW+04} 
but it is unknown at this time whether the hot medium is an extended Galactic
corona or one that pervades the Local Group.
If the gas density were uniform this would raise $N_{crit}$ appreciably
as well.
However, if pressures are slightly larger than previous estimates,
the gas may be inhomogeneous with pockets of cold \ion{H}{1} at 
densities $\sim 100$ times larger than the warm neutral medium.
If so, $N_{crit}$ could again be of order 
$2 \times 10^{19}$~${\rm cm}^{-2}$ where collisional ionization is present
(and much less where it is not).

\subsection{Implications for Lyman Limit Systems}

The distribution function of central column density $N_{HI}$ from 
the northern hemisphere Leiden-Dwingeloo Survey (LDS) of \citet{dHBB02b} 
and the southern hemisphere Parkes survey (HIPASS) of \citet{P+02}
is shown in Fig. 11, along with that for our mini-HVCs.
We do not know yet what selection effects apply to the mini-HVCs;
the areas searched with the necessary column density sensitivity
may well be favorable one.
So we do not know how many more mini-HVCs there might be on the whole
sky.
For illustrative purposes, in Fig. 11 we have simply multiplied the 
mini-HVC distribution by a factor of 100 to bring it into agreement
with that for CHVC for $N_{HI}$ near $10^{19}$~${\rm cm}^{-2}$.
If mini-HVCs were distributed around the sky similar to CHVCs, the 
factor would be much larger.
If the factor of 100 proves to be of the right order of magnitude,
Fig. 11 suggests that low-column density CHVC have been undercounted by
the surveys and that there are many more in total.

There has been some controversy as to what extent Lyman Limit Systems are composed of 
small starless individual gas clouds.
CHVC, and their lower column density extensions, are candidates for this
purpose, especially if it turns out that low column density CHVC have been undercounted.

\section{Conclusions and Summary}

We have presented evidence of 14 very small, low column density 
\ion{H}{1} clouds which we call ``mini-HVC.''
These have peak neutral hydrogen column densities (averaged over the 
Arecibo $3\farcm2$ beam) $\lesssim 10^{19}~{\rm cm}^{-2}$ and 
\ion{H}{1} diameters to our minimum detectable column density 
$\lesssim 10^{18}~{\rm cm}^{-2}$ in the range $9\arcmin$ to $35\arcmin$.
Both column density and diameter are considerably smaller than most Compact High Velocity Clouds
(CHVC), but the mini-HVC fit smoothly onto the low column density tail of the CHVC in
a plot of column density vs. diameter.
Although our mapping undersamples the mini-HVCs, we find near-constant 
column densities around the observed peak.
This suggests that shock activity cannot be prevalent and that there
is no transition from the warm neutral to cold neutral phase in these clouds.
The sensitivity of our detection survey for mini-HVC is considerable better than that for the CHVC catalogs, 
but only a very small fraction of the sky was covered.
Consequently, we cannot compare number densities for the two surveys reliably.
However, it is possible that the presently available CHVC catalogs are appreciably incomplete for \ion{H}{1}
column densities $\lesssim 10^{19}~{\rm cm}^{-2}$.

For CHVC recent surveys have given unexpectedly large values of H$\alpha$ intensity 
\citep{TWMHR02, PBV+03}.
This has raised questions as to whether ionization by extragalactic UV photons is greatly
supplemented by Galactic UV and/or by collisions with a hot intra-Local Group
medium \citep{BB02}.
The ratio of ionized to neutral hydrogen is therefore somewhat uncertain for CHVC.
Because of the smaller \ion{H}{1} column densities, this ratio is even more uncertain, 
and possibly quite large, for mini-HVC.

There remain two chief unknowns about mini-HVC:  their distance from the Milky Way and
their distribution around the sky.
To determine the latter would require mapping much larger areas of the sky with the
Arecibo L-band Feed Array (ALFA), but to much higher sensitivity than is planned for the
ALFA consortium surveys.
A distribution on the sky similar to that of CHVC would suggest that the mini-HVC
are simply the low column density tail of the CHVC distribution.
Although there continues to be much controversy on distances
\citep{BBdH02, MP03}, typical CHVC may lie at distances of order 50-100 kpc
and mini-HVCs at distances of order 200 kpc.
One other possibility is a relationship to the Magellanic Stream, except for 
one mini-HVC which is nearly a hemisphere away.
Recently, some gas clouds have been detected by \citet{TBW+04} which seem to lie within 
about 50 kpc of M31.
This raises the possibility that the mini-HVC are CHVC analogs within $\sim 50$ to 100 kpc
from M31 rather than in a comparable volume around the Milky Way.
However, in this case the M31 CHVC would have to have smaller column densities on average than 
the Milky Way CHVC.

\acknowledgments

\clearpage

\begin{deluxetable}{c c c c c c c c c}
\tabletypesize{\footnotesize}
\rotate
\tablewidth{0pt}
\tableheadfrac{0.2}
\tablenum{1}
\tablecaption{Results of Mapping}
\tablecolumns{9}
\tablehead{
\colhead{Object} & \colhead{RA(2000)} & \colhead{Dec(2000)} & 
\colhead{Peak Flux} & \colhead{Diameters} & \colhead{P.A.} &
\colhead{Peak Column} & \colhead{$V_{sys}$} & \colhead{$\Delta V_{50}$}  \\
\colhead{} & \colhead{(hhmmss.s)} & \colhead{(ddmmss)} & 
\colhead{(Jy km ${\rm s}^{-1}$)} & \colhead{(arcmin)} & \colhead{(deg)} &
\colhead{($10^{18} {\rm cm}^{-2}$)} & \colhead{(km~${\rm s}^{-1}$)} & 
\colhead{(km~${\rm s}^{-1}$)} }
\startdata
WvW~413 & 222135.6 & 254709 & 0.515 & $42 \times 35$ & 10 & 14.5 & $-425$ & 27 \\
WvW~479 & 012427.8 & 254003 & 2.370 & $35 \times 24$ & 40 & 66.9 & $-344$ & 21 \\
WvW479min1 & 012409.0 & 252403 & 0.164 & $>15 \times 6$ & \nodata & 
4.6 & $-248$ & 27 \\
WvW479min2 & 012741.7 & 252715 & 0.212 & $>19 \times 10$ & \nodata & 
6.0 & $-245$ & 29 \\
LMPU025-163 & 005305.9 & 291710 & 0.080 & $35 \times 13$ & 70 & 2.2 & $-175$ & 22 \\
LMPU025-291 & 005434.9 & 292158 & 0.276 & $16 \times 13$ & 160 & 7.8 & $-289$ & 25 \\
LMPU025-335 & 005331.3 & 292334 & 0.149 & $22 \times 19$ & 40 & 4.2 & $-335$ & 20 \\
LMPU025-395 & 005123.8 & 292022 & 0.571 & $>22 \times 16$ & 0 & 16.1 & $-395$ & 21 \\
LMPU025 OFF-173 & 005944.0 & 293758 & 0.114 & $>22 \times 11$ & 60 & 
3.2 & $-173$ & 22 \\
LMPU025 OFF-293 & 005956.7 & 291710 & 0.129 & $>9 \times 6$ & \nodata & 
3.6 & $-293$ & 27 \\
LMPU236+242 & 123955.5 & 093441 & 0.068 & $19 \times 6$ & 10 & 1.9 & $+242$ & 27 \\
LMPU369-210 & 213849.5 & 023140 & 0.289 & $>35 \times 19$ & 60 & 8.2 & $-198$ & 49 \\
LMPU369 OFF-260 & 214545.0 & 030204 & 0.157 & $>14 \times 10$ & 30 & 
4.4 & $-268$ & 32 \\
LMPU387-230 & 225751.6 & 072649 & 0.169 & $22 \times 7$ & 60 & 4.8 & $-228$ & 23 \\
LMPU387-275 & 225740.4 & 075401 & 0.184 & $22 \times 11$ & 0 & 5.2 & $-275$ & 34 \\
LMPU387-297 & 225751.6 & 072337 & 0.318 & $>19 \times 16$ & 20 & 9.0 & $-297$ & 27 \\
\enddata
\end{deluxetable}

\begin{figure}
\figurenum{1}
\plotone{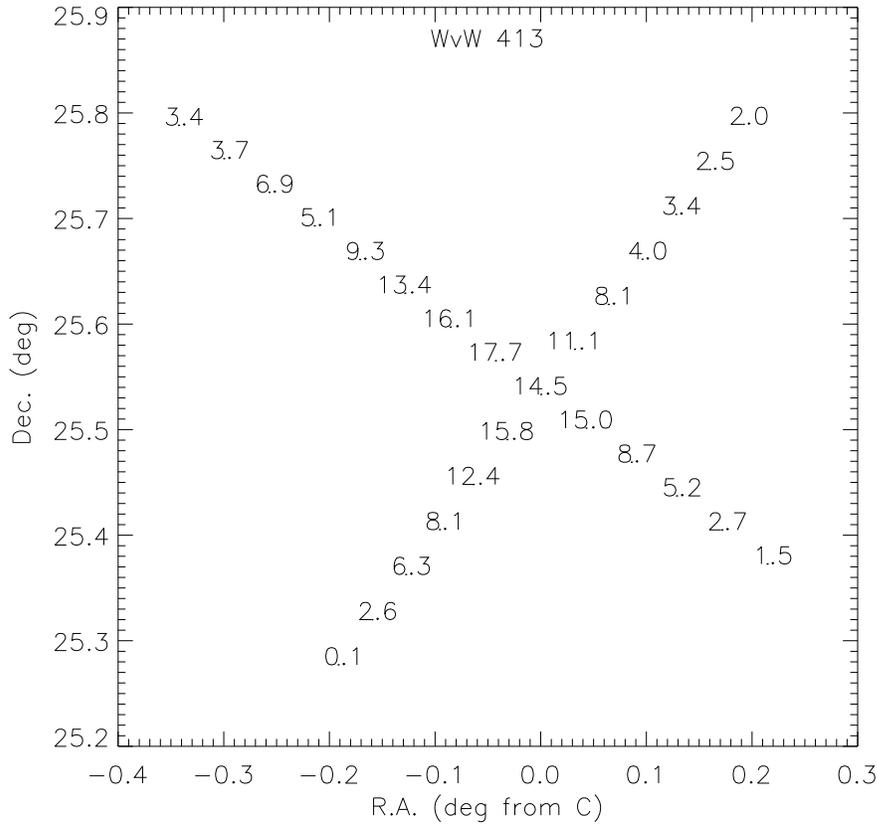}
\caption{
Map of observed column densities at velocities near $-425$~km~${\rm s}^{-1}$
along two diameters of WvW 413.
Negative R.A. offsets from 22:19:15.1 (B1950) are eastward.
The numbers (positioned with the observed point
at the center of the bottom edge of each entry) indicate detected emission,
in units of $10^{18}{\rm cm}^{-2}$, at the HVC velocities.}
\end{figure}

\begin{figure}
\figurenum{2}
\plotone{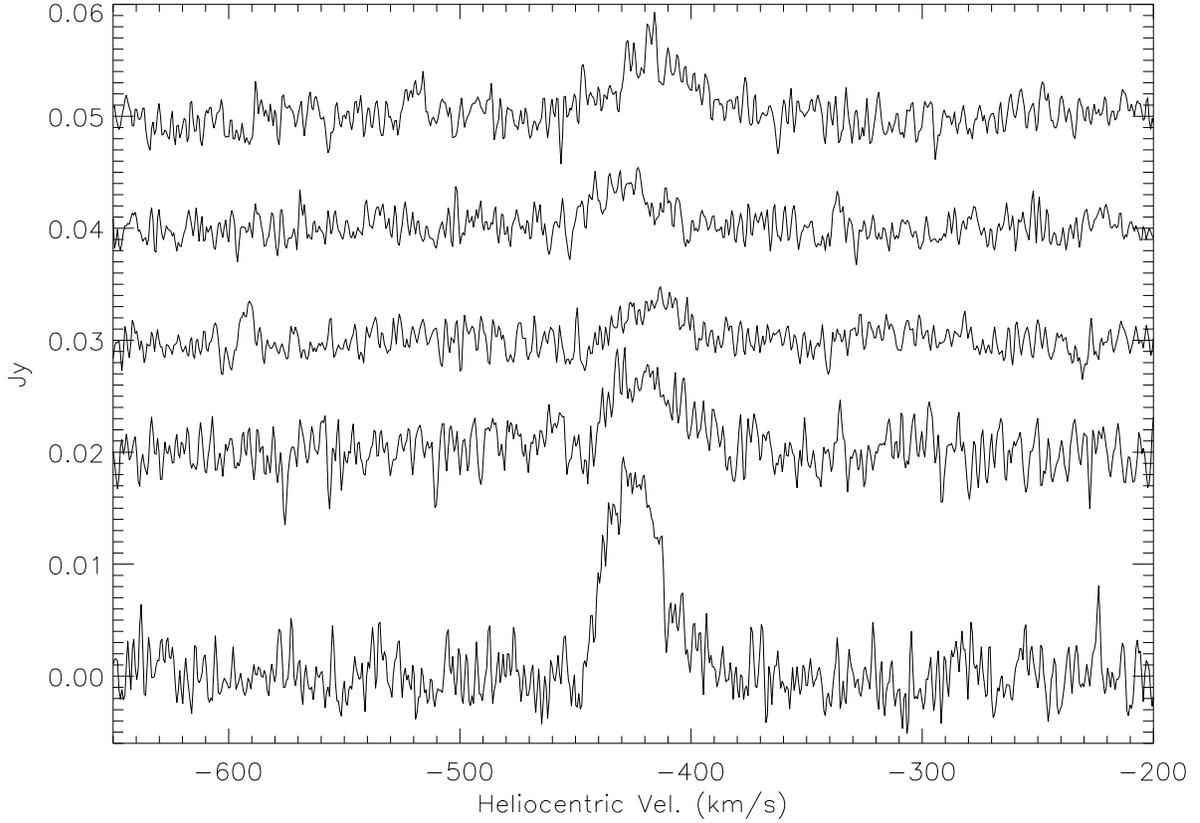}
\caption{
Spectra from the highest column density point at the core of WvW 413 and
from points near the outer edges of each diameter.
The core spectrum is plotted with no offset.
The spectrum from the point $19\fm2$ toward Galactic East is offset by 0.02 Jy, 
that from $12\fm8$ toward Galactic West by 0.03 Jy, that from $12\fm8$ toward 
Galactic North by 0.04 Jy,that from $12\fm8$ toward Galactic South by 0.05 Jy.}
\end{figure}

\begin{figure}
\figurenum{3}
\plotone{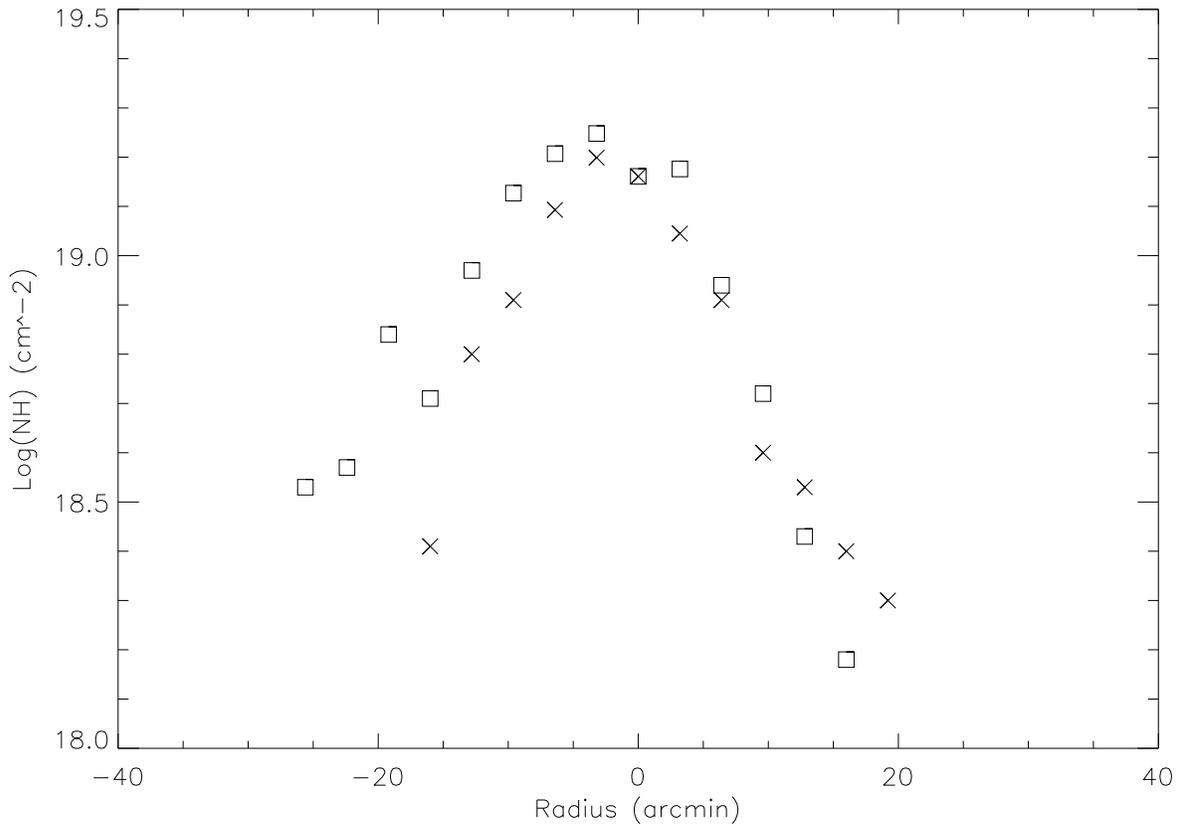}
\caption{
Logarithm of observed neutral column density at mapped points in WvW 413
as a function of radius along the Galactic E-W axis (squares) and 
Galactic N-S axis (exes).
Galactic East and South are plotted as negative radii.}
\end{figure}

\begin{figure}
\figurenum{4}
\plotone{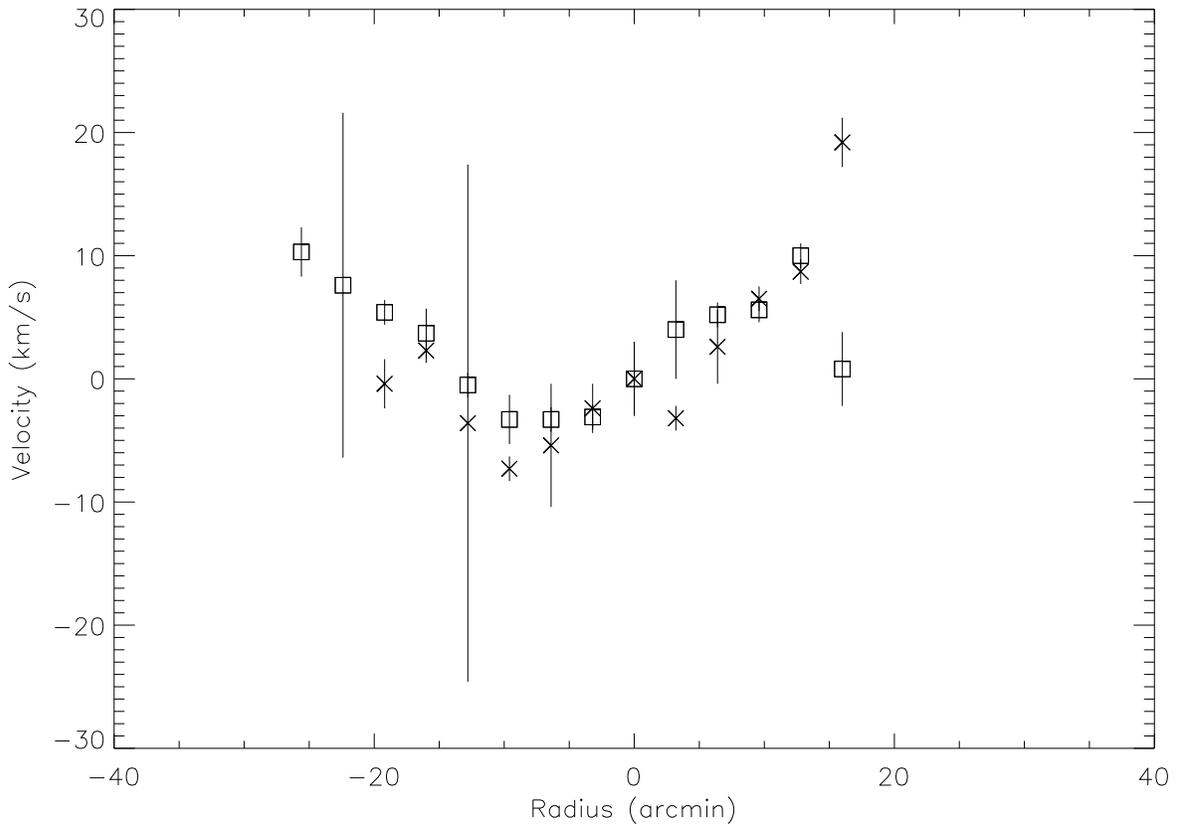}
\caption{
A ``rotation curve'' for WvW~413, showing systemic velocity (centroid of gaussian fit
to the profile) as a function
of radius along the Galactic E-W axis (squares) and Galactic N-S axis
(exes).
East and South are taken to be negative.
Velocities are with respect to the systemic velocity obtained for the
center profile, $-425.0$~km~${\rm s}^{-2}$.
}
\end{figure}

\begin{figure}
\figurenum{5}
\plotone{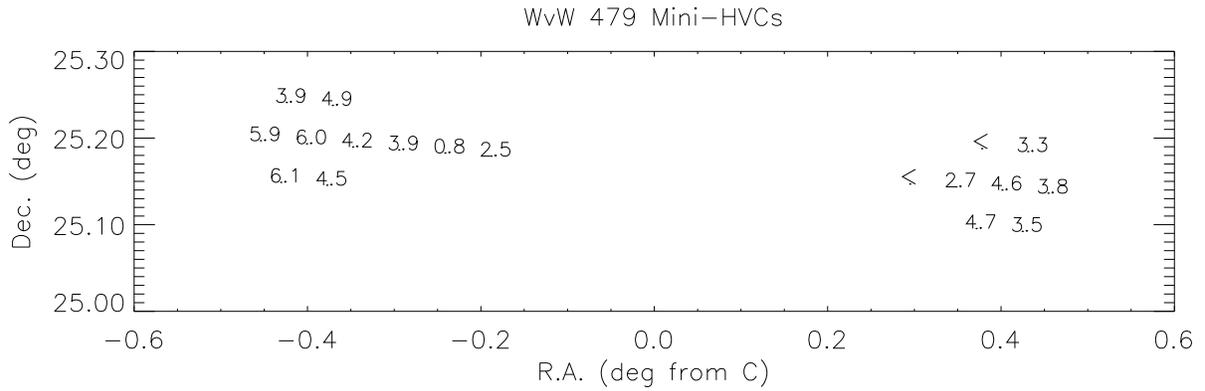}
\caption{
Map of observed column densities at velocities near $-248$~km~${\rm s}^{-1}$
showing the two mini-HVCs superimposed on the edge of WvW 479.
Negative R.A. offsets from 01:23:12.35 (B1950) are eastward.
The core of WvW 479 itself has R.A. 01:21:44.8, Dec. 25:24:24 (B1950) and
heliocentric velocity $-344$~km~${\rm s}^{-1}$.
The numbers (positioned with the observed point
at the center of the bottom edge of each entry) indicate detected emission,
in units of $10^{18}{\rm cm}^{-2}$, at the mini-HVC velocities.
Non-detections are indicated by $<$.}
\end{figure}

\begin{figure}
\figurenum{6}
\plottwo{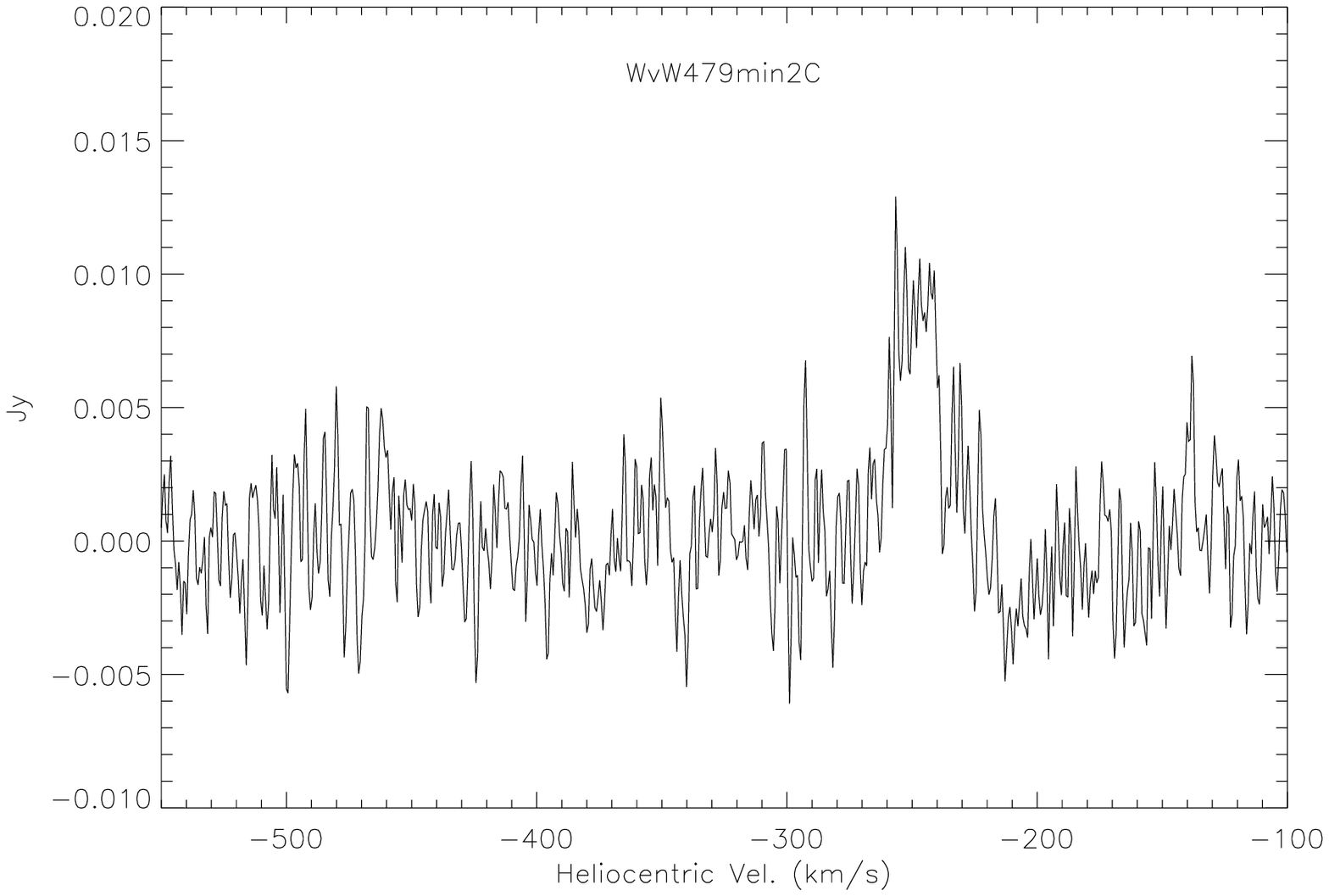}{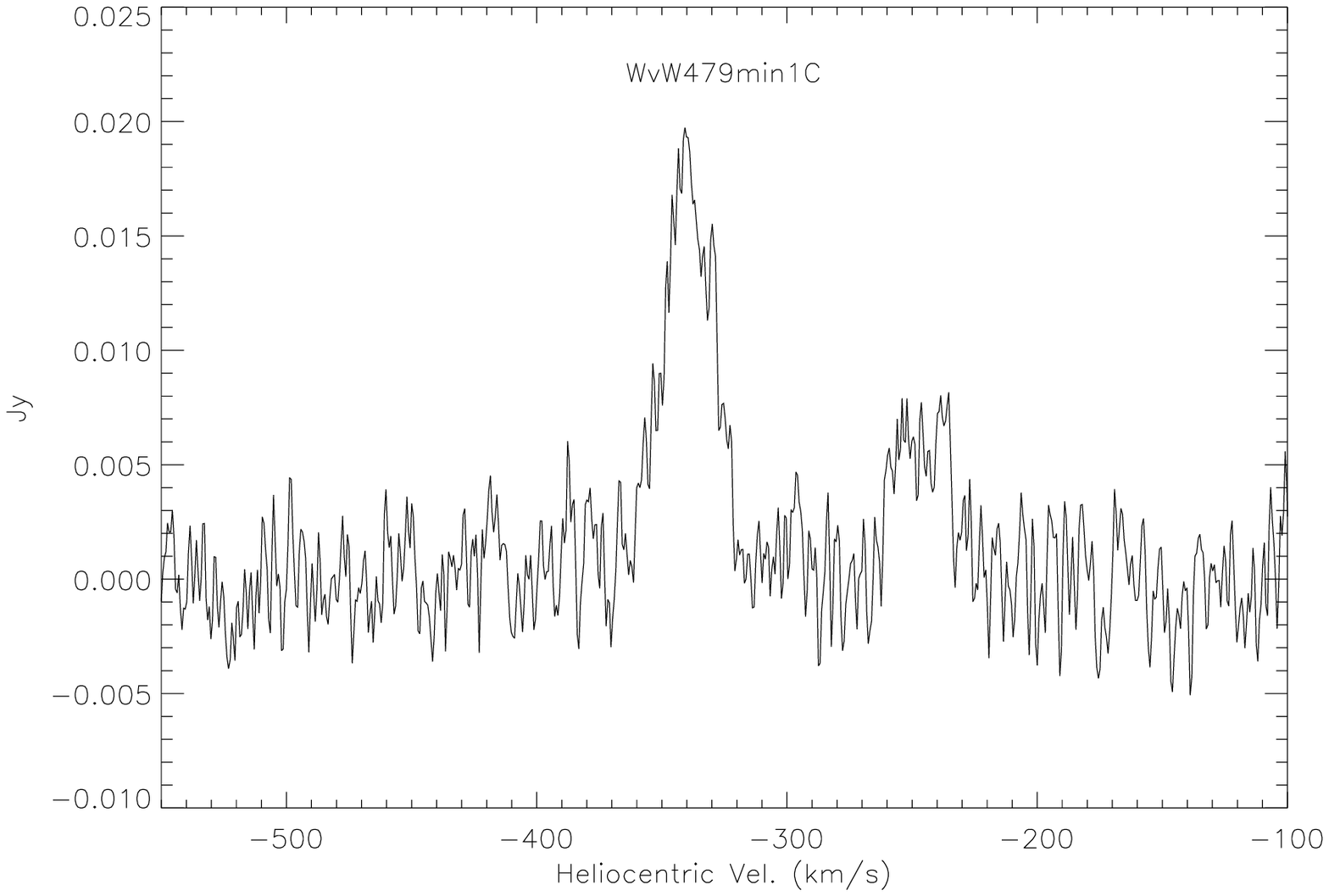}
\caption{
Spectra from the highest column density points, ($-0.40106$, 25.1933) and 
(0.40106, 25.1400) respectively, in the eastern (left) and western
(right) mini-HVCs superimposed on WvW 479.
The right-hand spectrum shows some emission from the edge of WvW 479 itself 
at $-338$~km~${\rm s}^{-1}$ in addition to that from the mini-HVC at
at $-248$~km~${\rm s}^{-1}$.}
\end{figure}

\begin{figure}
\figurenum{7a}
\plotone{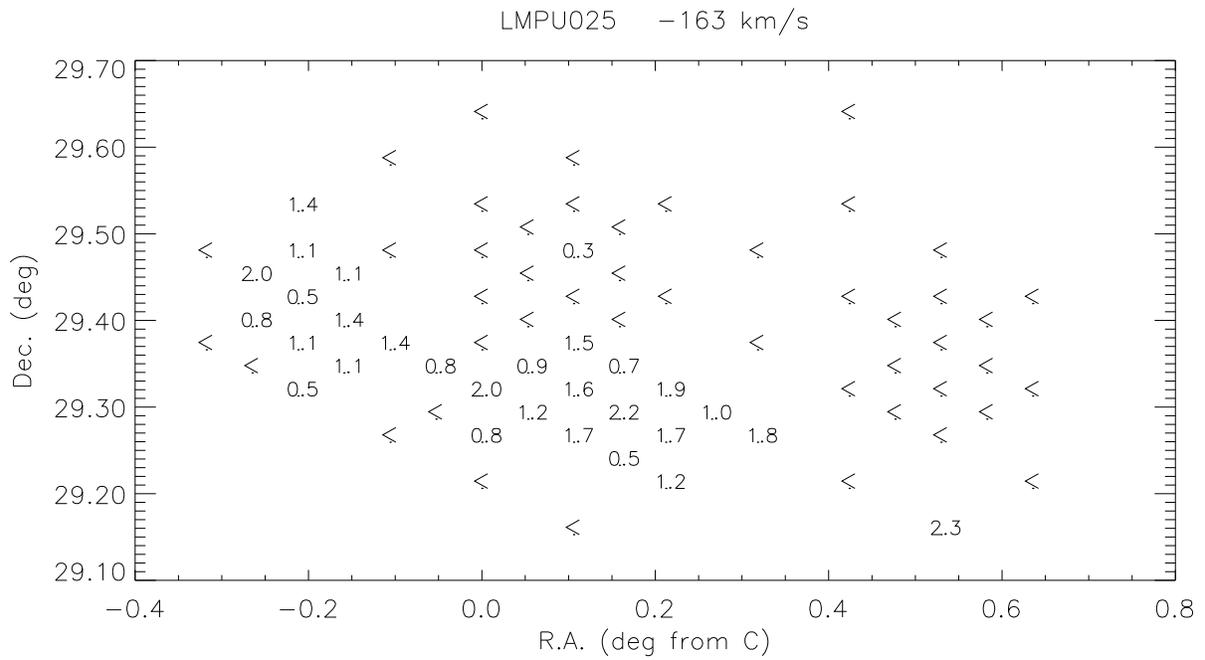}
\caption{
Map of observed column densities at velocities near $-163$~km~${\rm s}^{-1}$
around LMPU 025.
Negative R.A. offsets from 00:53:44.0 (J2000), the center of the LMPU beam, 
are eastward.
Numbers (positioned as in Fig. 5) 
indicate detected emission, in units of $10^{18}{\rm cm}^{-2}$, at this 
velocity.
Non-detections are indicated by $<$.}
\end{figure}

\begin{figure}
\figurenum{7b}
\plotone{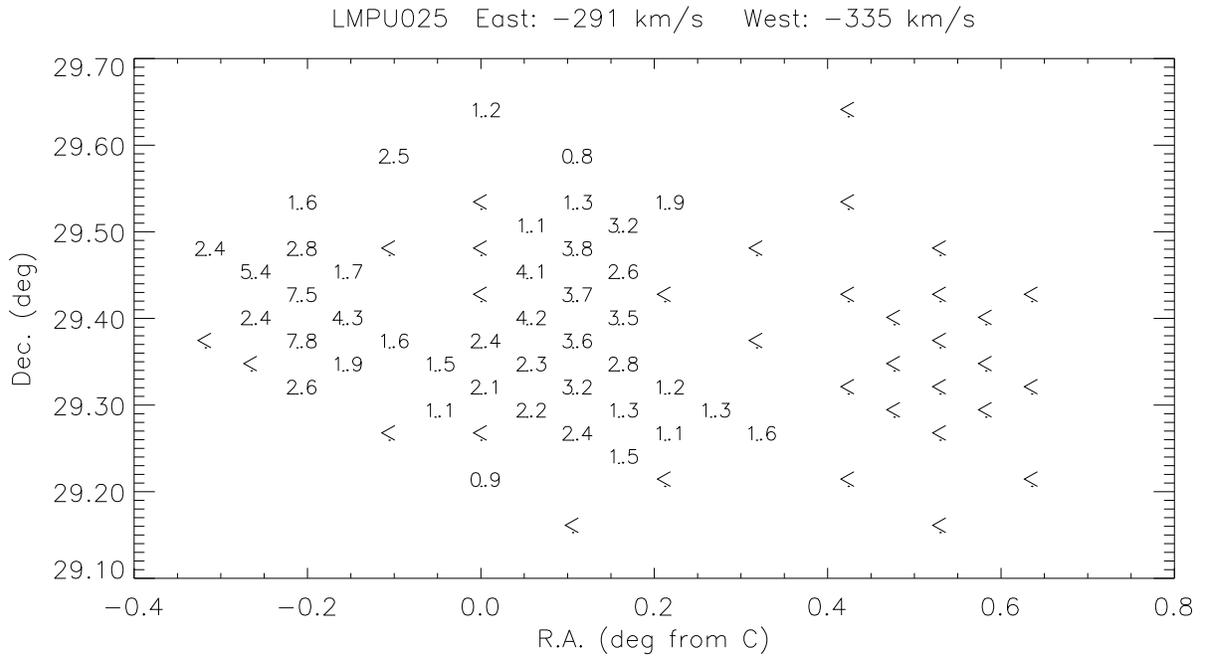}
\caption{
Velocities near $-300$~km~${\rm s}^{-1}$ around LMPU 025.
Map center and symbols are the same as Fig. 7a.
Detections east (negative offsets) of offset $-0.12$ along with those
at ($-0.1$,29.58) and (0.0,29.63) are at heliocentric velocities
near $-291$~km~${\rm s}^{-1}$; those westward from that point are at
$-335$~km~${\rm s}^{-1}$.}
\end{figure}

\begin{figure}
\figurenum{7c}
\plotone{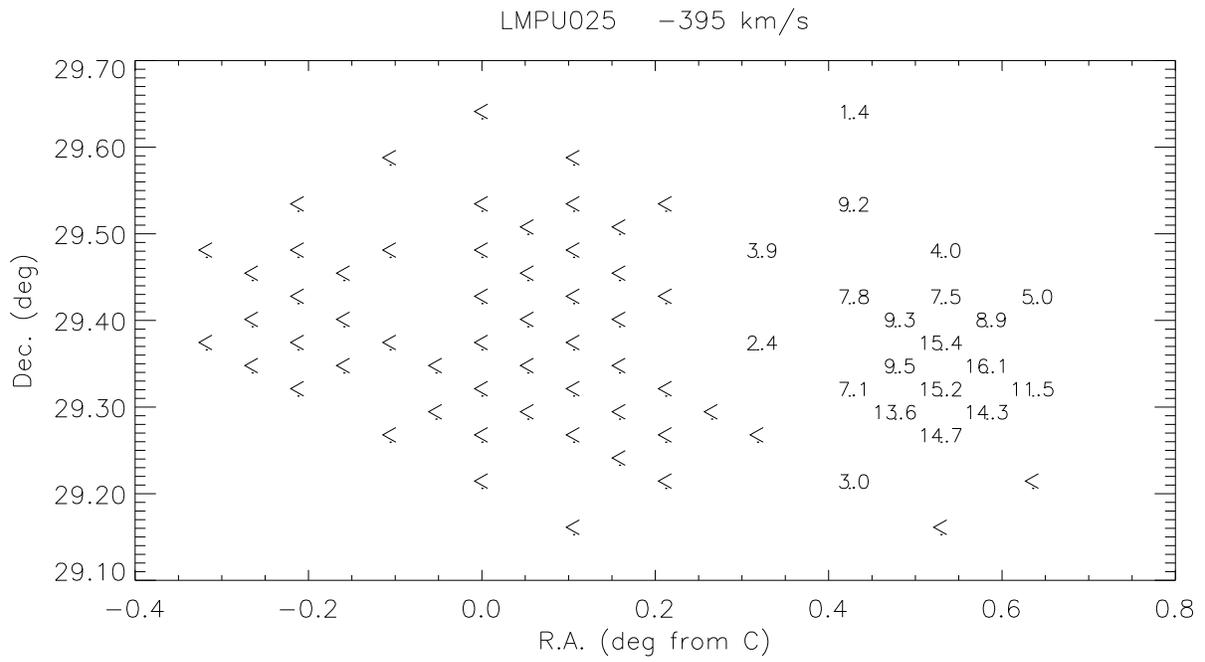}
\caption{
Velocities near $-395$~km~${\rm s}^{-1}$
just west of LMPU 025.
Map center and symbols are the same as Fig. 7a.}
\end{figure}

\begin{figure}
\figurenum{7d}
\plotone{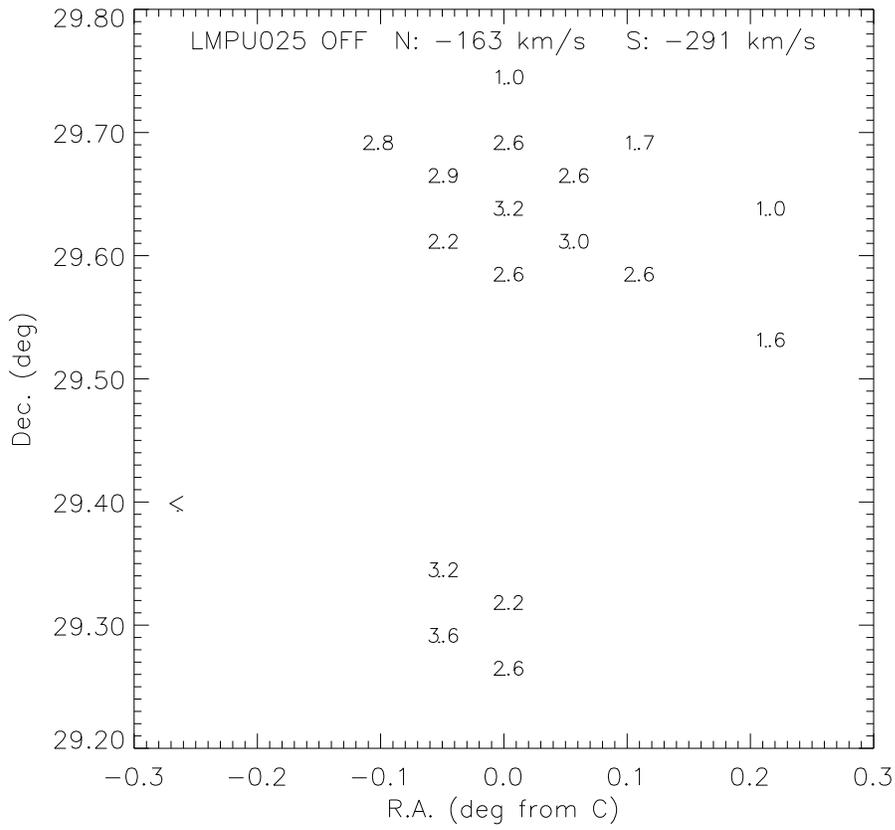}
\caption{
Velocities near $-173$~km~${\rm s}^{-1}$
(northern cloud) and $-293$~km~${\rm s}^{-1}$ (southern cloud) in the
reference beams for LMPU 025.
Negative R.A. offsets from 00:59:44.0 (J2000) are eastward.
Symbols are coded as in Fig. 7a.}
\end{figure}

\begin{figure}
\figurenum{7e}
\plotone{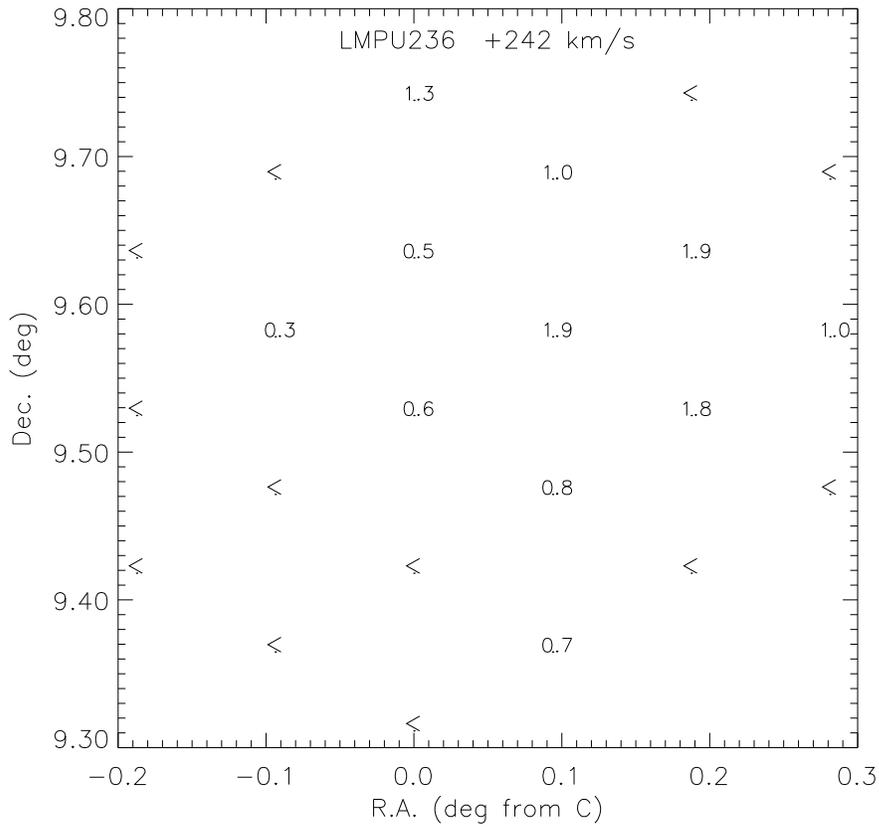}
\caption{
Velocities near $242$~km~${\rm s}^{-1}$
within the field of LMPU 236.
Negative R.A. offsets from 12:40:18.0 (J2000) are eastward.
Symbols are coded as in Fig. 7a.}
\end{figure}

\begin{figure}
\figurenum{7f}
\plotone{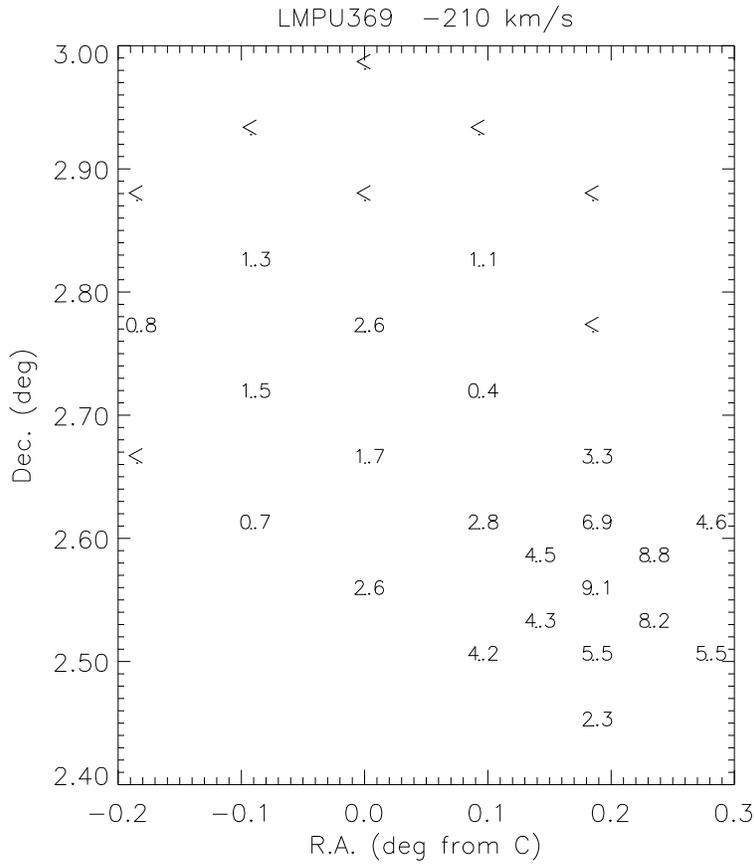}
\caption{
Velocities near $-210$~km~${\rm s}^{-1}$ around the field of LMPU 369.
Negative R.A. offsets from 21:39:45.0 (J2000) are eastward.
Symbols are coded as in Fig. 7a.}
\end{figure}

\begin{figure}
\figurenum{7g}
\plotone{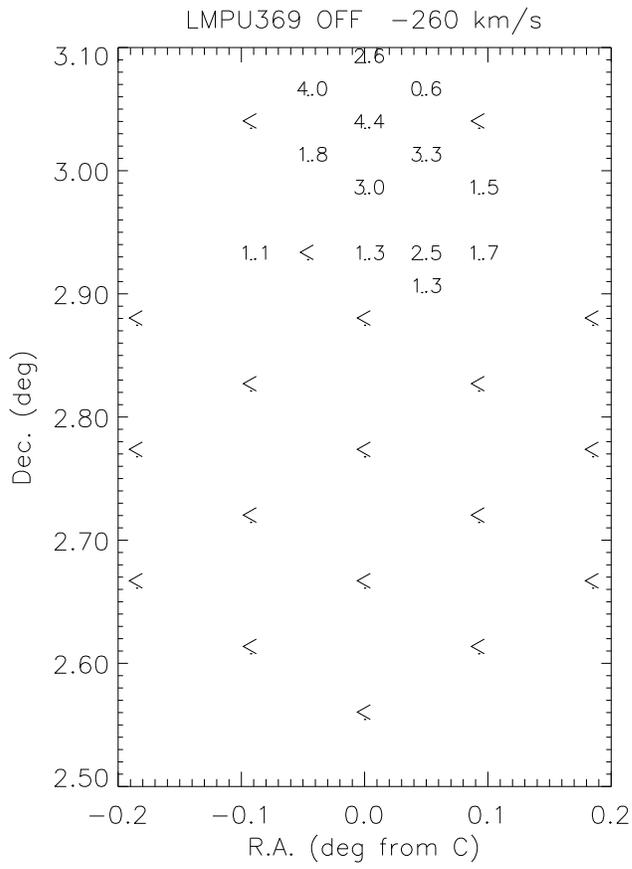}
\caption{
Velocities near $-260$~km~${\rm s}^{-1}$ in the field of reference beams for LMPU 369.
Negative R.A. offsets from 21:45:45.0 (J2000) are eastward.
Symbols are coded as in Fig. 7a.}
\end{figure}

\begin{figure}
\figurenum{7h}
\plotone{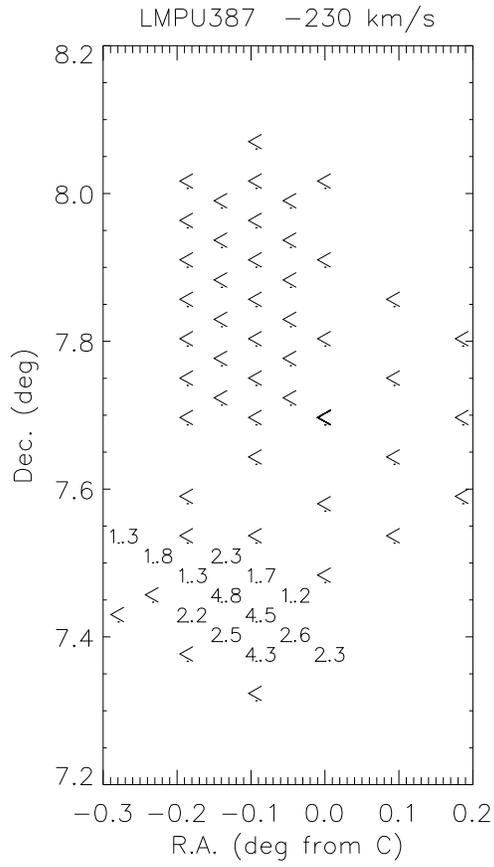}
\caption{
Velocities near $-230$~km~${\rm s}^{-1}$ around the field of LMPU 387.
Negative R.A. offsets from 22:57:18.0 (J2000) are eastward.
Symbols are coded as in Fig. 7a.}
\end{figure}

\begin{figure}
\figurenum{7i}
\plotone{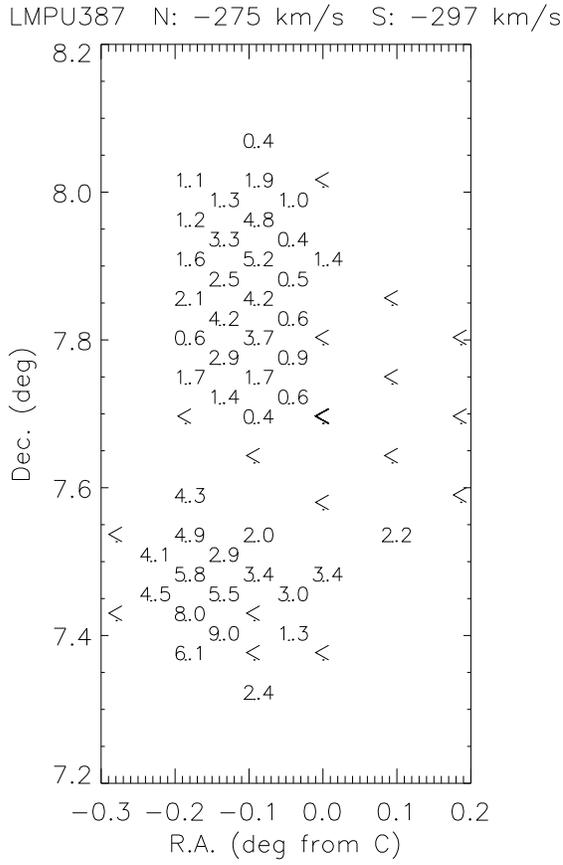}
\caption{
Velocities near $-275$~km~${\rm s}^{-1}$ (northern cloud) and
$-290$~km~${\rm s}^{-1}$ (southern cloud) around the field of LMPU 387.
Field center is the same as in Fig. 7h, and symbols are coded as in Fig. 7a.}
\end{figure}

\clearpage

\begin{figure}
\figurenum{8a}
\plotone{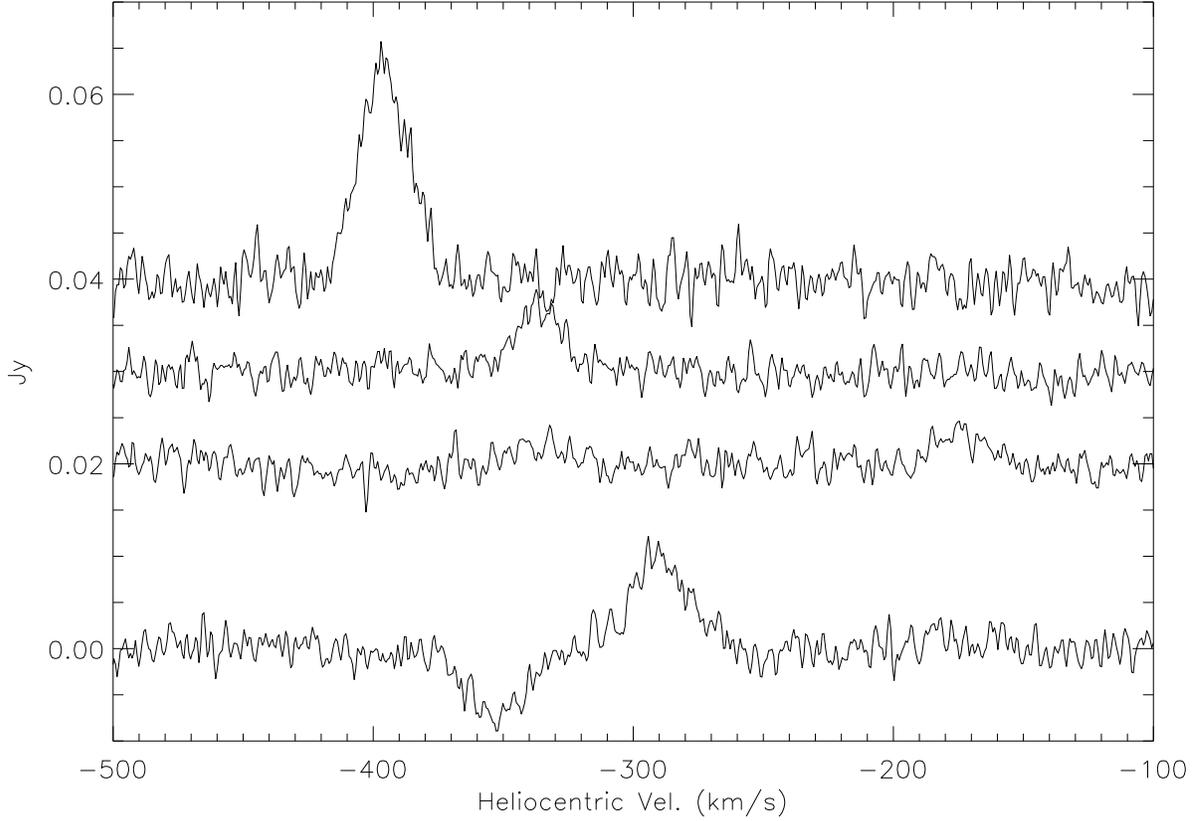}
\caption{
Spectra (no offset) from the highest column density point (0.1588, 29.2861) in the 
$-163$~km~${\rm s}^{-1}$ mini-HVC around LMPU 025 with weak emission from
the $-335$~km~${\rm s}^{-1}$ mini-HVC as well, (offset by 0.02 Jy) from a point
(-0.2121, 29.3661) showing emission from the $-291$~km~${\rm s}^{-1}$ mini-HVC,
(offset by 0.03 Jy) from the highest column density point (0.0529, 29.3928) in the 
$-335$~km~${\rm s}^{-1}$ mini-HVC, and (offset by 0.04 Jy) from a point
(0.5296, 29.3128) showing the highest column density emission from the 
$-395$~km~${\rm s}^{-1}$ mini-HVC.
The apparent negative signal is due to the HVC WvW 466 in the reference beam.}
\end{figure}

\begin{figure}
\figurenum{8b}
\plotone{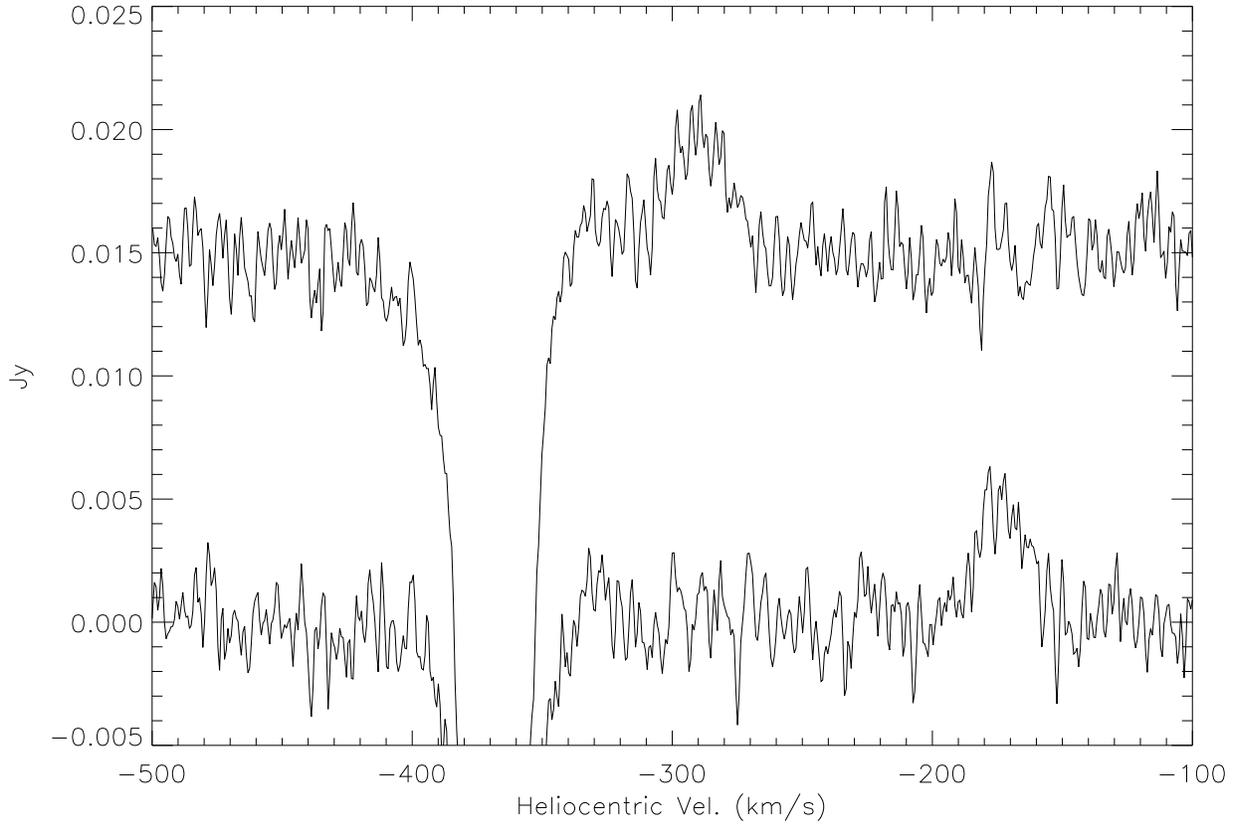}
\caption{
Spectra (no offset) from the highest column density point (0.0, 29.6328) in the 
$-173$~km~${\rm s}^{-1}$ mini-HVC in the reference beams for LMPU 025 
and (offset by 0.015 Jy) from the highest column density point (-0.0529, 29.2861) 
in the $-293$~km~${\rm s}^{-1}$ mini-HVC.
The apparent negative signal is due to the HVC WvW 466 in the reference 
beams ($6^{m}$ still further east).}
\end{figure}

\begin{figure}
\figurenum{8c}
\plotone{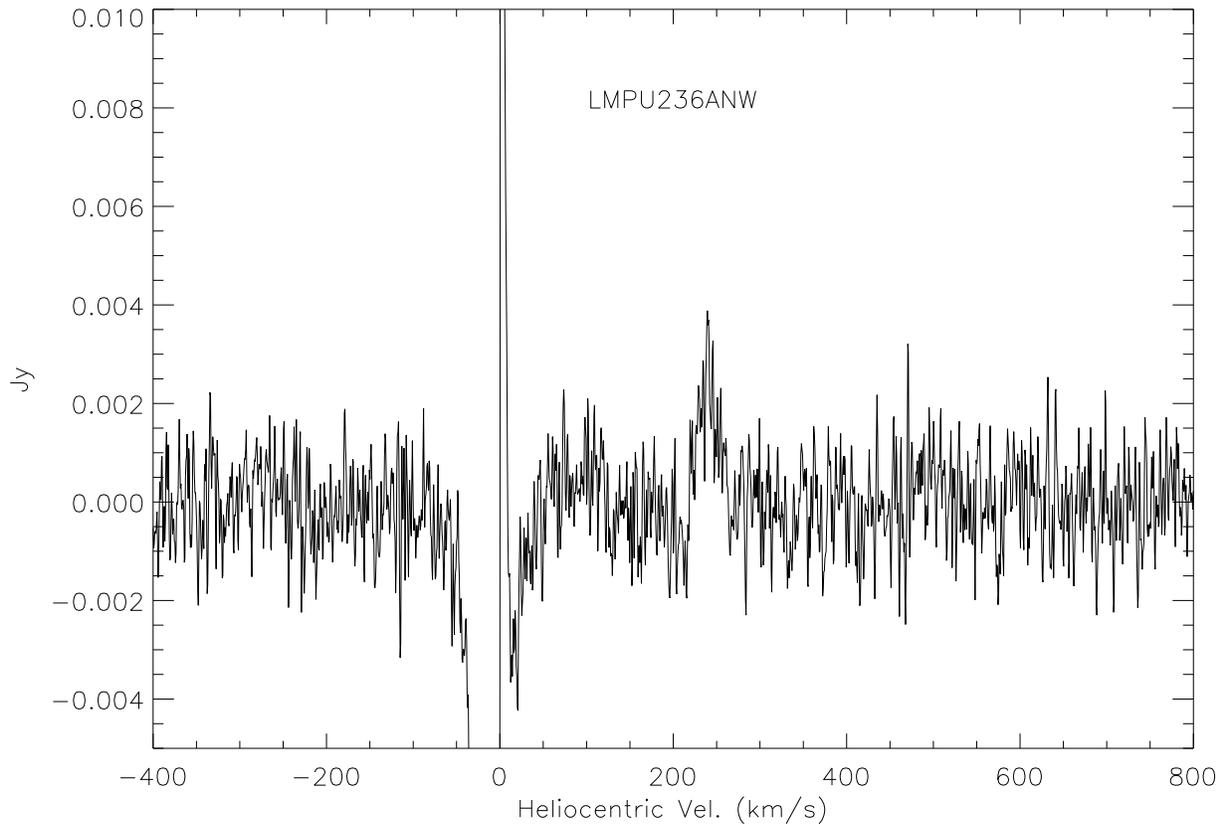}
\caption{
Spectrum from the highest column density point (0.0938, 9.5781) in the 
LMPU 236 mini-HVC.
Incompletely cancelled Galactic \ion{H}{1} emission is evident in the 
heliocentric velocity range $-100$-$+50$~km~${\rm s}^{-1}$.}
\end{figure}

\begin{figure}
\figurenum{8d}
\plotone{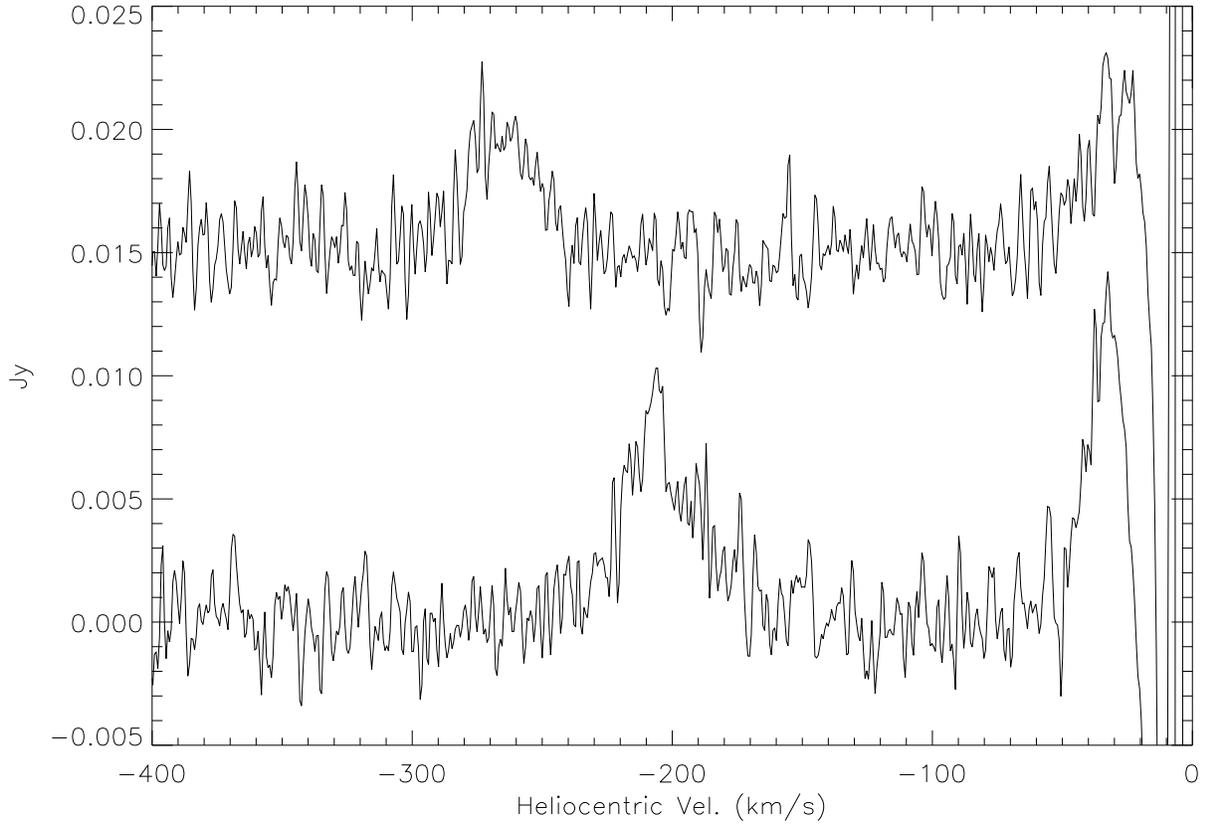}
\caption{
Spectra from the highest column density points, (0.2312, 2.5278) in the 
LMPU 369 mini-HVC (no offset) and (0.0000, 3.0344) in the reference beam mini-HVC
(offset by 0.015 Jy).
Incompletely cancelled Galactic \ion{H}{1} emission is evident at 
heliocentric velocities $> -50$~km~${\rm s}^{-1}$.}
\end{figure}

\begin{figure}
\figurenum{8e}
\plotone{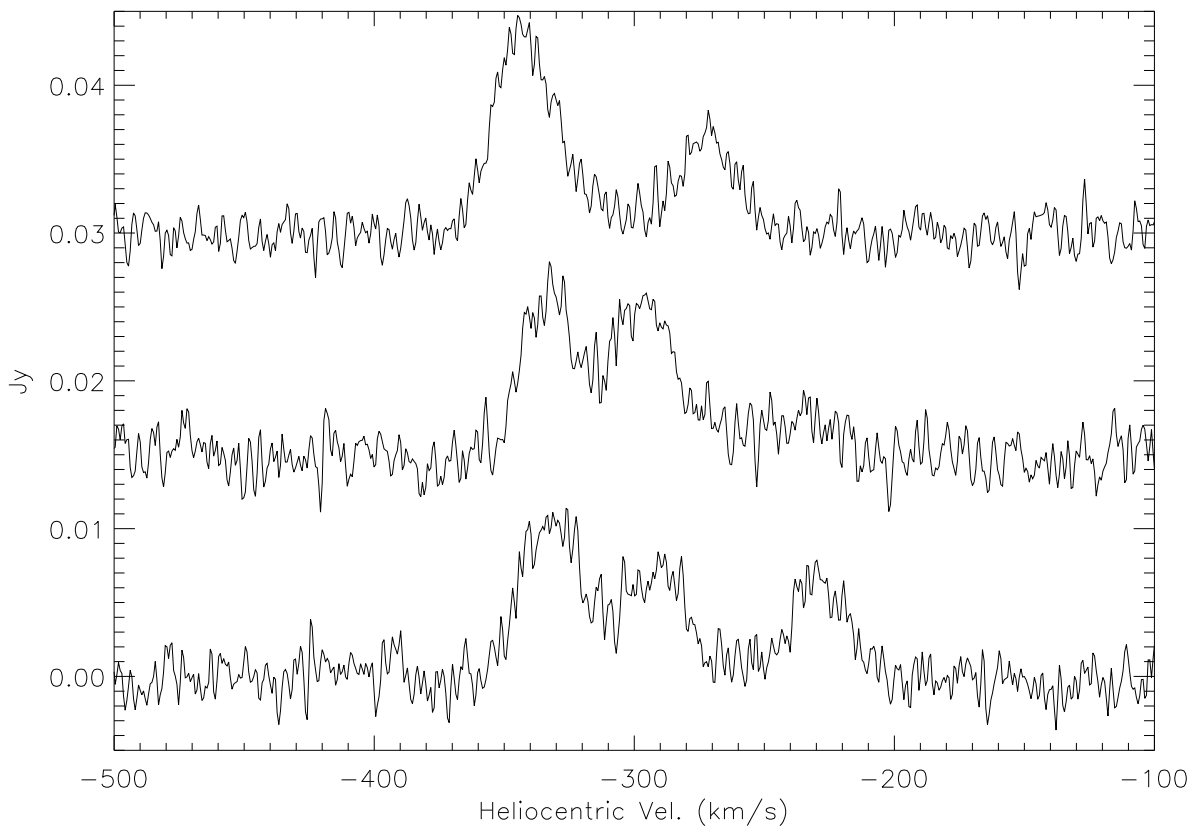}
\caption{
Spectra from the highest column density points, ($-0.1400$, 7.4469), 
($-0.1400$, 7.3936) and ($-0.1400$, 7.8197) respectively, in the (no offset) 
$-230$~km~${\rm s}^{-1}$, (offset by 0.015 Jy) $-290$~km~${\rm s}^{-1}$ and 
(offset by 0.03 Jy) $-275$~km~${\rm s}^{-1}$ mini-HVCs around LMPU 387.
The no-offset spectrum shows some emission from the $-290$~km~${\rm s}^{-1}$ 
mini-HVC as well, and all three show emission from the Magellanic Stream at 
$-336$~km~${\rm s}^{-1}$.}
\end{figure}

\begin{figure}
\figurenum{9}
\plottwo{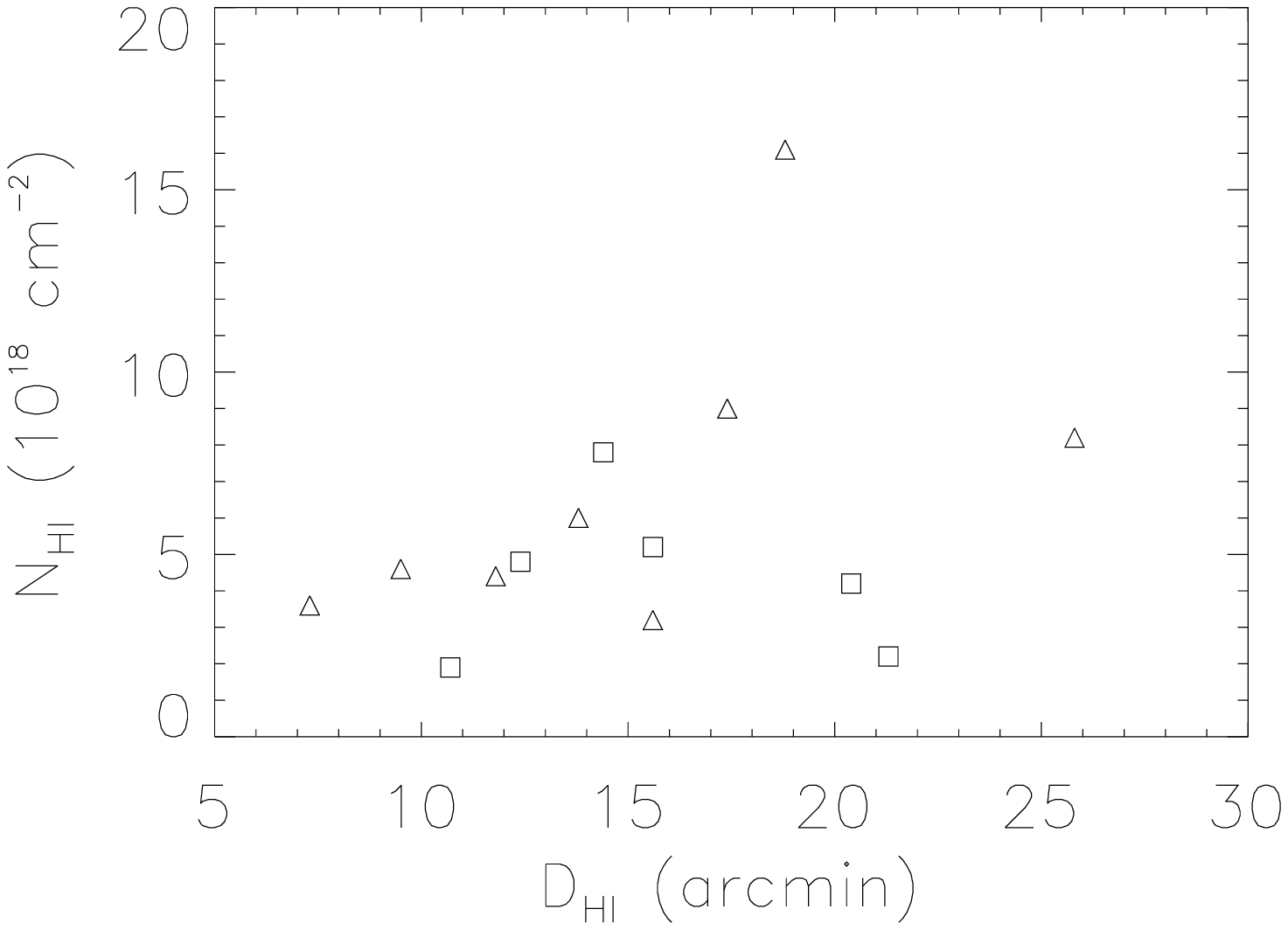}{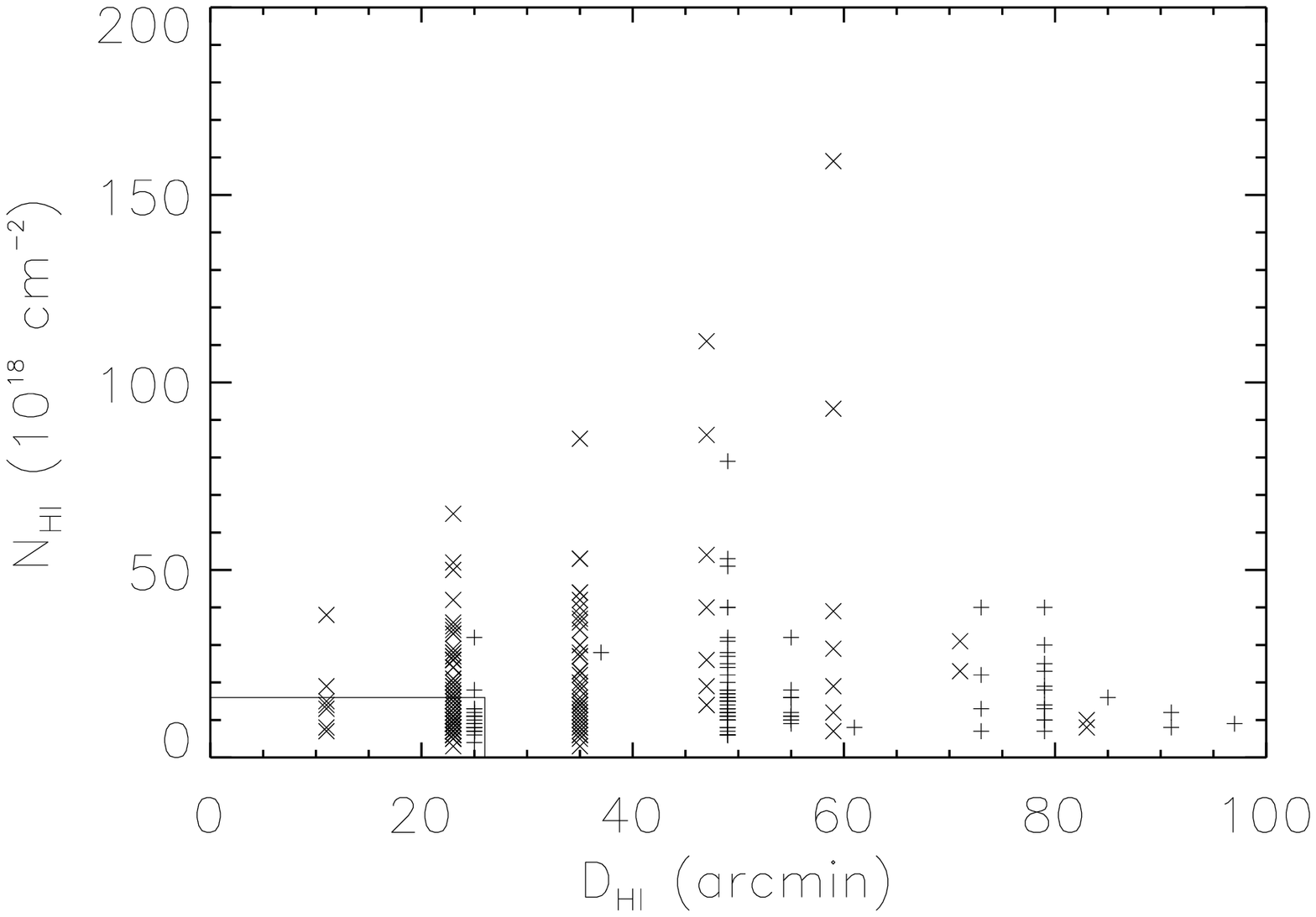}
\caption{
Peak column density $N_{HI}$ averaged over the observing beam vs. \ion{H}{1} diameter $D_{HI}$ (geometric mean of major and minor axes)
for our mini-HVC alone (left) and for CHVC (right).
In the left-hand figure, squares indicate mini-HVCs for which our detection limit
was reached on all sides, while triangles indicate those for which $D_{HI}$
is a lower limit.
In the right-hand figure, CHVCs from LDS are shown
as plus signs, those from HIPASS as exes.
The small box in the lower left corner of the CHVC figure indicates the upper limits
of $D_{HI}$ and $N_{HI}$ for the mini-HVC.
Both LDS and HIPASS used beams considerably larger than Arecibo.}
\end{figure}

\begin{figure}
\figurenum{10}
\plotone{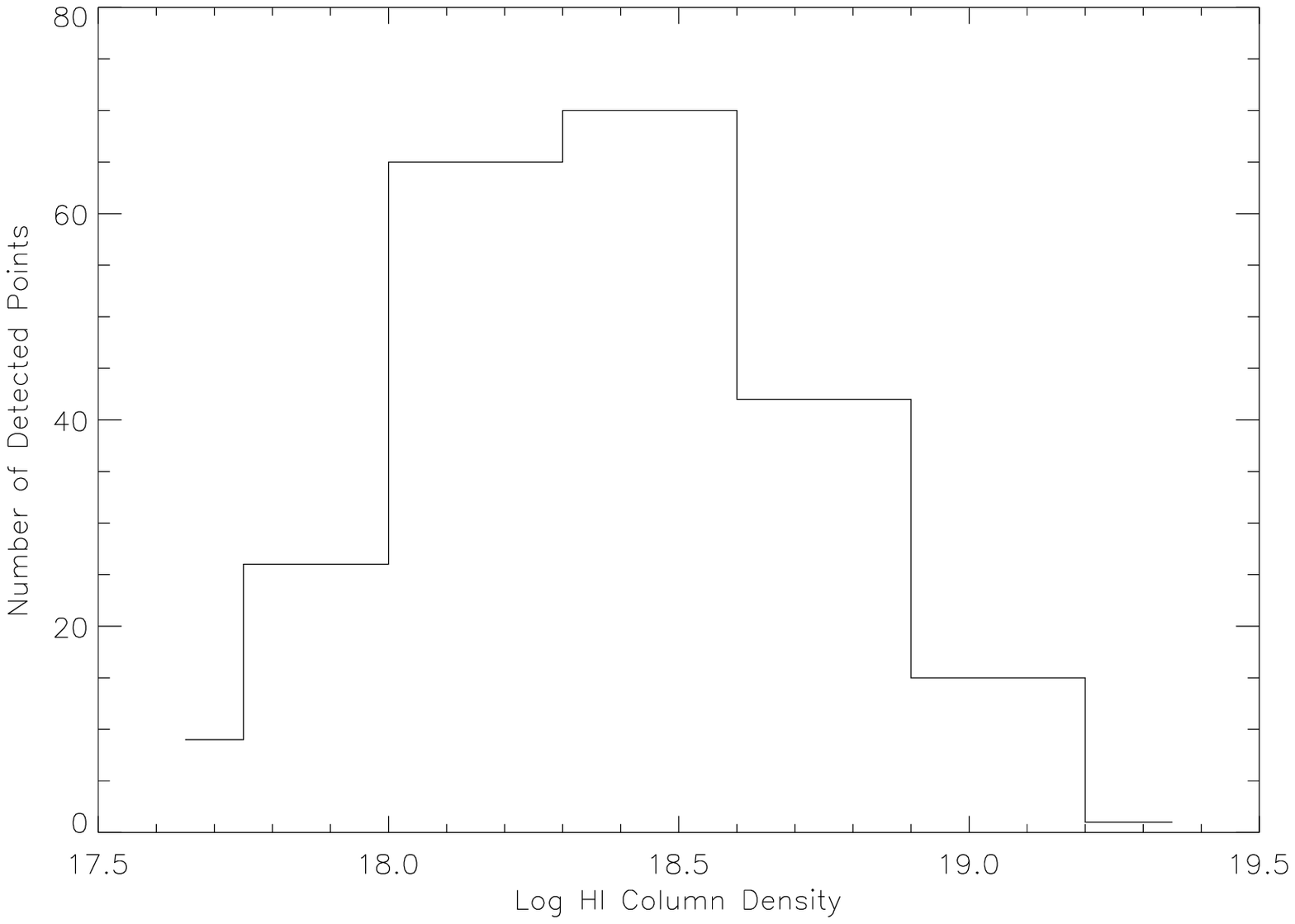}
\caption{
Histogram of logarithmic column density for all mini-HVC beam positions at which 
emission was detected.}
\end{figure}

\begin{figure}
\figurenum{11}
\plotone{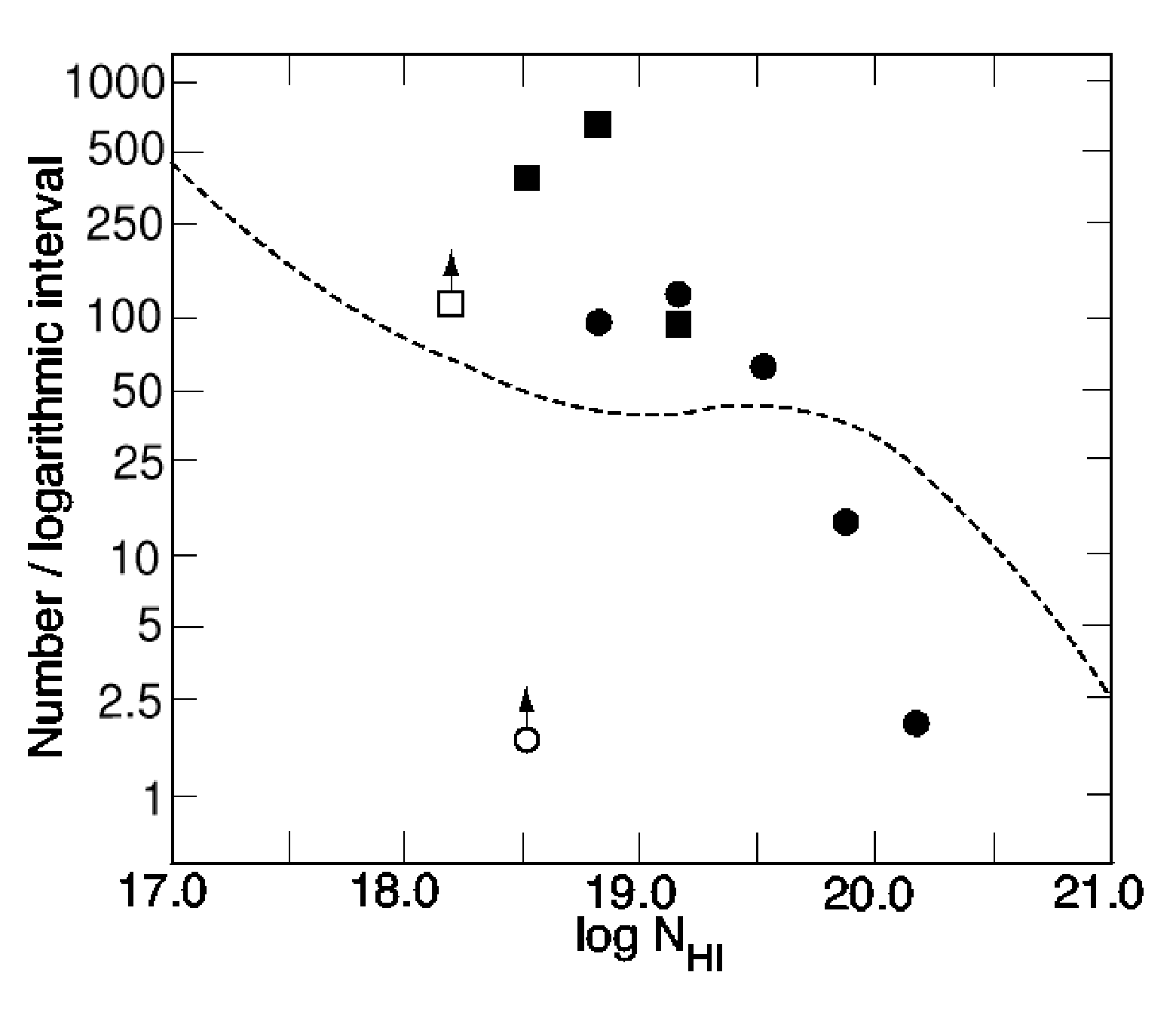}
\caption{
Distribution of CHVC and mini-HVC in logarithmic column density.
Circles represent CHVC from the whole sky.
Squares represent mini-HVC scaled upward as discussed in the text.
Arrows indicate incompleteness.
The dotted curve is a schematic drawn from results in \citet{CB02} 
and \citet{CSB01}.}
\end{figure}


\begin{thebibliography}{}

\bibitem[Blitz et al. (1999)]{BSTHB99} Blitz, L., Spergel, D., Teuben, P., 
Hartmann, D., \& Burton, W. B.  1999, \apj, 514, 818

\bibitem[Braun \& Burton (1999)]{BB99} Braun, R., \& Burton, W. B.
1999, \aap, 341, 437

\bibitem[Braun \& Thilker (2004)]{BT04} Braun, R., \& Thilker, D. A.
2004, \aap, 417, 421

\bibitem[Burton, Braun \& Chengalur (2001)]{BBC01} Burton, W. B., Braun, R.,
\& Chengalur, J.  2001, \aap, 369, 616

\bibitem[Burstein \& Blumenthal (2002)]{BB02} Burstein, D., \& 
Blumenthal, G.  2002, \apj, 574, L17

\bibitem[Burton, Braun \& de Heij (2002)]{BBdH02} Burton, W. B.,
Braun, R., \& de Heij, V.  2002, to appear in {\em High Velocity Clouds},
eds. van Woerden, Wakker, Schwarz \& de Boer (astro-ph/0206359)

\bibitem[Corbelli \& Bandiera (2002)]{CB02} Corbelli, E., \& Bandiera, R.
2002, \apj, 567, 712

\bibitem[Corbelli, Salpeter \& Bandiera (2001)]{CSB01} Corbelli, E., 
Salpeter, E. E., \& Bandiera, R.  2001, \apj, 550, 26

\bibitem[Davies et al. (2002)]{DSD+02} Davies, J., Sabatini, S., Davies, L.,
Linder, S., Roberts, S., Smith, R., \& Evans, Rh.  2002, \mnras, 336, 155

\bibitem[de Heij, Braun \& Burton (2002a)]{dHBB02a} de Heij, V., Braun, R., 
\& Burton, W. B.  2002a, \aap, 391, 159

\bibitem[de Heij, Braun \& Burton (2002b)]{dHBB02b} de Heij, V., Braun, R., 
\& Burton, W. B.  2002b, \aap, 392, 417

\bibitem[Fox et al. (2004)]{FSW+04} Fox, A. J., Savage, B. D., Richter, P.,
Sembach, K. R., \& Tripp, T. M.  2004, \apj, 602, 738

\bibitem[Hoffman et al. (2003)]{HBSC03} Hoffman, G. L., Brosch, N., Salpeter, E. E.,
\& Carle, N. J.  2003, \aj, 126, 2774

\bibitem[Hoffman, Salpeter \& Carle (2001)]{HSC01} Hoffman, G. L., 
Salpeter, E. E., \& Carle, N. J.  2001, \aj, 122, 2428

\bibitem[Hoffman, Salpeter \& Pocceschi (2002)]{HSP02} Hoffman, G. L., 
Salpeter, E. E., \& Pocceschi, M. G.  2002, \apj, 576, 232

\bibitem[Hopp, Schulte-Ladbeck \& Kerp (2003)]{HSK03} Hopp, U., Schulte-Ladbeck, R. E.,
\& Kerp, J.  2003, \mnras, 339, 33

\bibitem[Lockman et al. (2002)]{LMPU02} Lockman, F. J., Murphy, E. M., 
Petty-Powell, S., \& Urick, V. J.  2002, \apjs, 140, 331

\bibitem[Maloney \& Putman (2003)]{MP03} Maloney, P. R., \& Putman, M. E.
2003, \apj, 589, 270

\bibitem[Murphy, Lockman, \& Savage (1995)]{MLS95} Murphy, E. M., 
Lockman, F. J., \& Savage, B. D.  1995, \apj, 447, 642

\bibitem[Putman et al. (2002)]{P+02} Putman, M. E., et al. 2002,
\aj, 123, 873  

\bibitem[Putman et al. (2003a)]{PBV+03} Putman, M. E., Bland-Hawthorn, J.,
Veilleux, S., Gibson, B. K., Freeman, K. C., \& Maloney, P. R.  2003a, \apj, 597, 948

\bibitem[Putman et al. (2003b)]{PSFGB03} Putman, M. E., Staveley-Smith, L., 
Freeman, K. C., Gibson, B. K., \& Barnes, D. G.  2003b, \apj, 586, 170

\bibitem[Robishaw, Simon \& Blitz (2002)]{RSB02} Robishaw, T., Simon, J. D., \& 
Blitz, L.  2002, \apj, 580, L129

\bibitem[Sembach et al. (2002)]{SGFP02} Sembach, K. R., Gibson, B. K., Fenner, Y., \& 
Putman, M. E.  2002, \apj, 572, 178

\bibitem[Simon \& Blitz (2002)]{SB02} Simon, J. D., \& Blitz, L.
2002, \apj, 574, 726

\bibitem[Stanimirovi\'{c} et al. (2002)]{SDKB02} Stanimirovi\'{c}, S.,
Dickey, J. M., Kr\'{c}o, M., \& Brooks, A. M.  2002, \apj, 576, 773

\bibitem[Sternberg, McKee \& Wolfire (2002)]{SMW02} Sternberg, A.,
McKee, C. F., \& Wolfire, M. G.  2002, \apjs, 143, 419 

\bibitem[Thilker et al. (2004)]{TBW+04} Thilker, D. A., Braun, R., Walterbos, R. A. M.,
Corbelli, E., Lockman, F. J., Murphy, E., \& Maddalena, R.  2004, \apj, 601, L39

\bibitem[Tufte et al. (2002)]{TWMHR02} Tufte, S. L., Wilson, J. D.,
Madsen, G. J., Haffner, L. M., \& Reynolds, R. J.  2002, \apj, 572, L153

\end{thebibliography}
\end{document}